\documentclass[aps
,prstab
,reprint
,longbibliography
,preprintnumbers
,showkeys
,amsfonts,amssymb,amsmath
,floatfix
]{revtex4-1}

\usepackage{bm}
\usepackage{dcolumn}
\usepackage{graphicx}
\usepackage[dvipsnames]{xcolor}
\usepackage[colorlinks=true, allcolors=NavyBlue]{hyperref}

\usepackage{mathptmx}
\usepackage[T1]{fontenc}

\newcommand{\q}[2]{\ensuremath{#1\ \mathrm{#2}}} % quantity with units
\newcommand{\code}[1]{\textsc{#1}} % computer code

\newcommand{\kth}{$k$th}
\newcommand{\kthtp}{$k$th-turn pulsing}
\newcommand{\seventhtp}{7th-turn pulsing}
\newcommand{\eighthtp}{8th-turn pulsing}
\newcommand{\tenthtp}{10th-turn pulsing}

\newcommand{\fma}{frequency-map analysis}

\DeclareFontFamily{U}{wncy}{}
\DeclareFontShape{U}{wncy}{m}{n}{<->wncyr10}{}
\DeclareSymbolFont{mcy}{U}{wncy}{m}{n}
\DeclareMathSymbol{\Sh}{\mathord}{mcy}{"58}

\newlength{\onethirdheight}
\newlength{\twothirdswidth}
\newlength{\halfwidth}
\newlength{\thirdwidth}
\newlength{\bsrtwidth}
\newlength{\fmawidth}
\newlength{\smallfmawidth}
\begin{document}

\title{Resonant and random excitations on the proton beam in the Large
  Hadron Collider for \\ active halo control with pulsed hollow
  electron lenses}

\author{Miriam Fitterer}
\author{Giulio Stancari}
\email[Corresponding author. E-mail: ]{stancari@fnal.gov}
\author{Alexander Valishev}
\affiliation{Fermi National Accelerator Laboratory, Batavia, Illinois, USA}

\author{Stefano Redaelli}
\author{Daniel Valuch}
\affiliation{CERN, Geneva, Switzerland}

\date{September 26, 2019}

\begin{abstract}
  We present the results of numerical simulations and experimental
  studies about the effects of resonant and random excitations on
  proton losses, emittances, and beam distributions in the Large
  Hadron Collider (LHC). In addition to shedding light on complex
  nonlinear effects, these studies are applied to the design of hollow
  electron lenses (HEL) for active beam halo control. In the
  High-Luminosity Large Hadron Collider (HL-LHC), a considerable
  amount of energy will be stored in the beam tails. To control and
  clean the beam halo, the installation of two hollow electron lenses,
  one per beam, is being considered. In standard electron-lens
  operation, a proton bunch sees the same electron current at every
  revolution. Pulsed electron beam operation (i.e., different currents
  for different turns) is also considered, because it can widen the
  range of achievable halo removal rates. For an axially symmetric
  electron beam, only protons in the halo are excited. If a residual
  field is present at the location of the beam core, these particles
  are exposed to time-dependent transverse kicks and to noise. We
  discuss the numerical simulations and the experiments conducted
  in~2016 and~2017 at injection energy in the LHC. The excitation
  patterns were generated by the transverse feedback and damping
  system, which acted as a flexible source of dipole kicks. Proton
  beam losses, emittances, and transverse distributions were recorded
  as a function of excitation patterns and strengths. The resonant
  excitations induced rich dynamical effects and nontrivial changes of
  the beam distributions, which, to our knowledge, have not previously
  been observed and studied in this detail. We conclude with a
  discussion of the tolerable and achievable residual fields and
  proposals for further studies.
\end{abstract}

%\pacs{}


\preprint{FERMILAB-PUB-18-084-AD-APC}

\maketitle

%\tableofcontents

\section{Introduction}
\label{sec:intro}

\begin{table*}
  \caption{Stored beam energy for a few examples of past, present and future
    colliders. New machines represents a leap in stored beam
    energy.}
  \label{tab:stored_energy}
  \begin{ruledtabular}
    \begin{tabular}{lccccc}
      Collider& Tevatron (protons) \cite{tevatron} & LHC 2016
                                                     \cite{Wenninger:IPAC:2017}
      & LHC nominal \cite{lhc_design} & HL-LHC \cite{hlcdr} & FCC \cite{Benedikt:JKPS:2016, Benedikt:IPAC:2018} \\
      \colrule
      Beam energy [TeV] & 0.98 & 6.5 & 7.0 & 7.0 & 50.0\\
      Number of bunches & 36 & 2220 & 2808 & 2748 & 10600 \\
      Number of particles per bunch & $2.90\times 10^{11}$ & $1.15\times 10^{11}$ & $1.15\times 10^{11}$ & $2.2\times 10^{11}$ & $1.0\times 10^{11}$\\
      Stored beam energy [MJ] & 1.6 & 265.9 & 362.2 & 678.0 & 8480 \\
    \end{tabular}
  \end{ruledtabular}
\end{table*}

\begin{figure*}
  \hspace*{\fill}
  \begin{minipage}[c]{0.45\textwidth}
    \includegraphics[width=\linewidth]{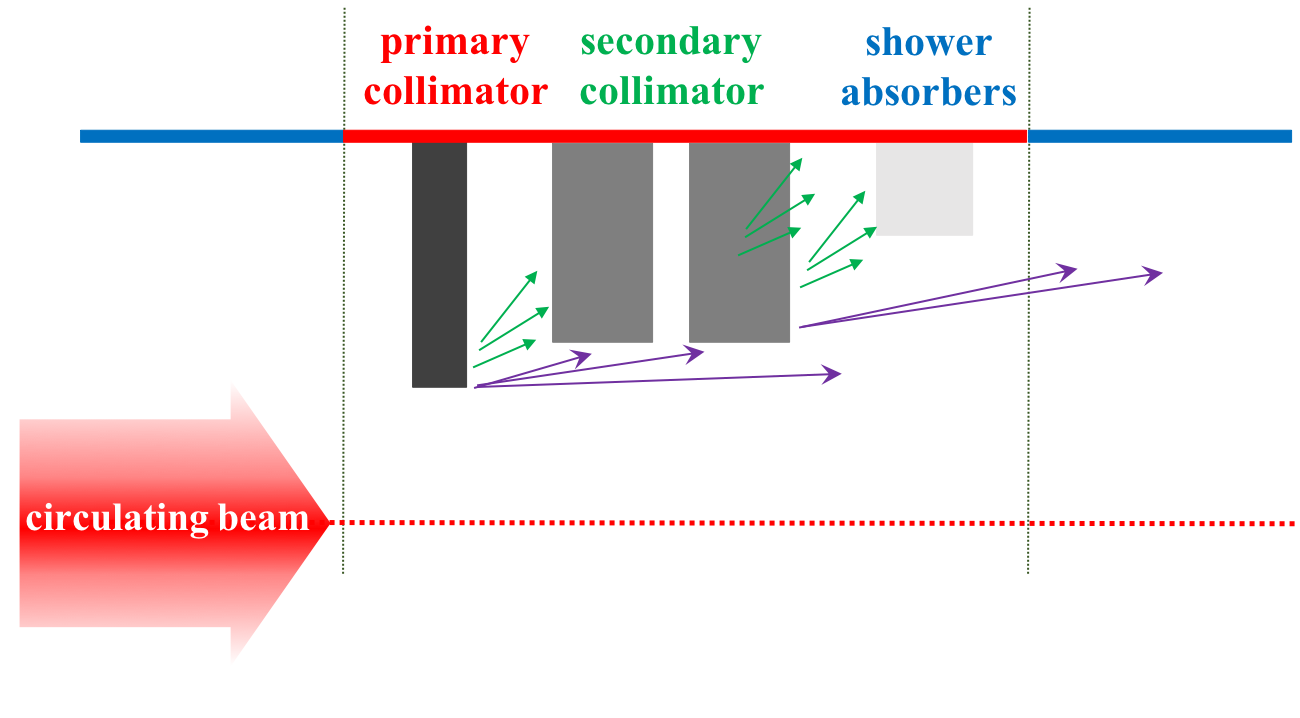}
    \includegraphics[width=\linewidth]{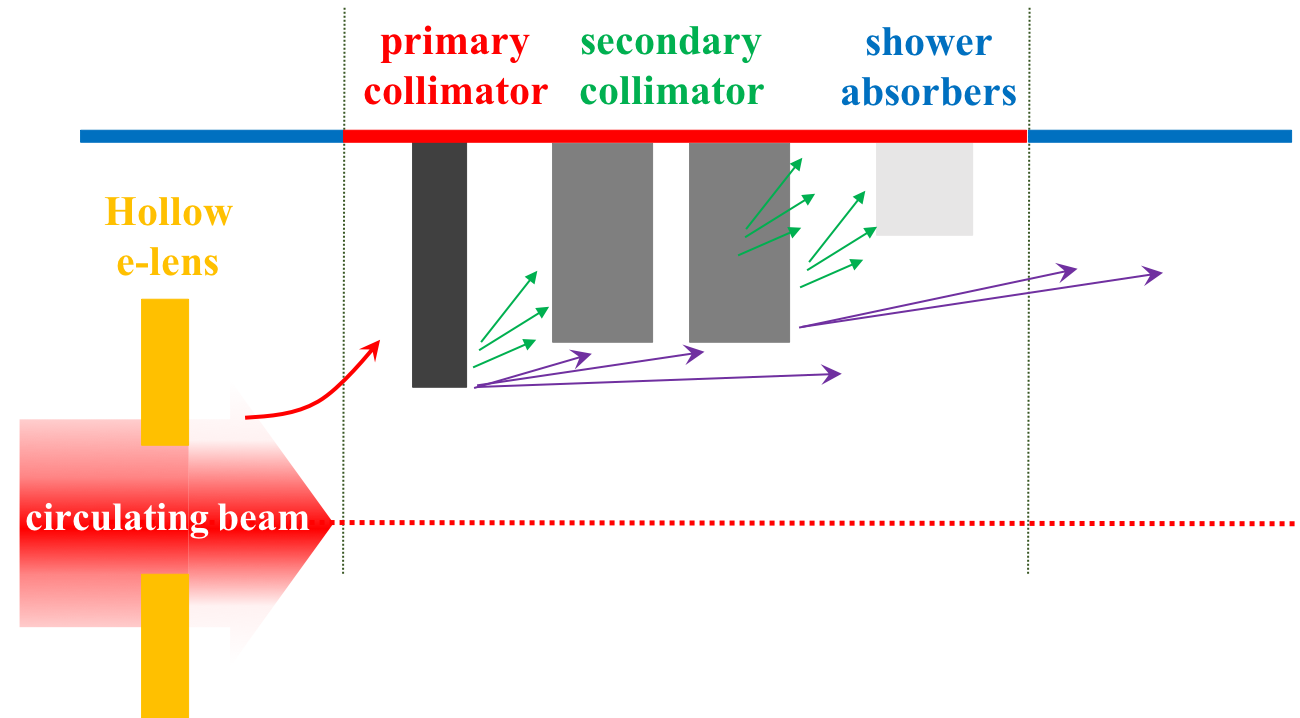}
  \end{minipage} \hfill
  \begin{minipage}[c]{0.4\textwidth}
    \includegraphics[width=\linewidth]{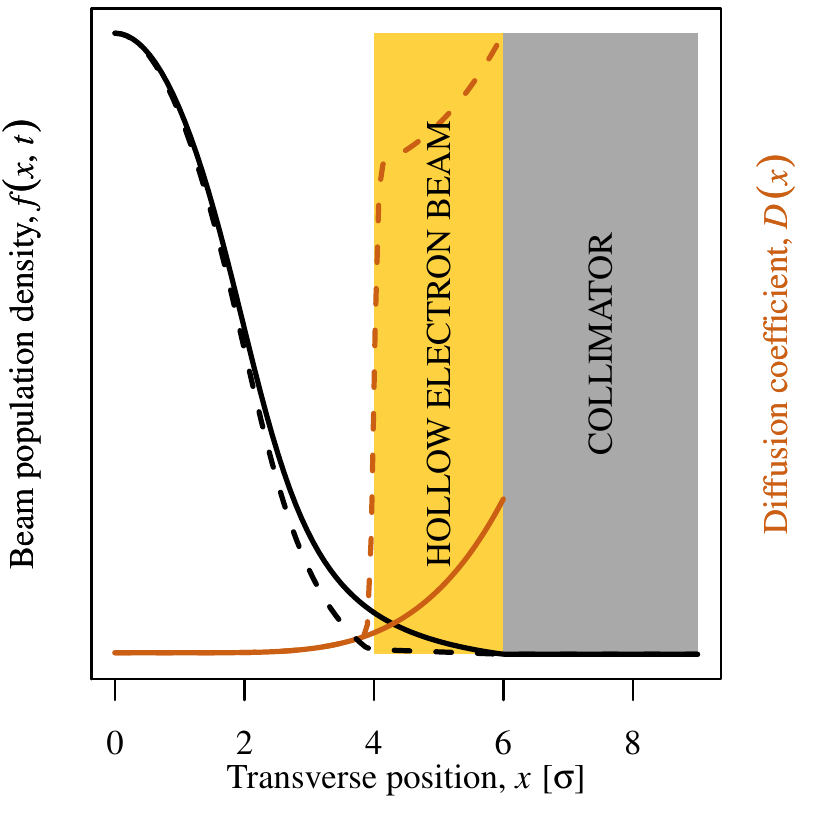}
  \end{minipage}
  \hspace*{\fill}
  \caption{Left: Sketch of passive halo control with a conventional
    collimation system (top) and active halo control, with the
    addition of a hollow electron lens (bottom). Right: Illustration
    of a simplified model of active diffusion enhancement in the
    transverse plane. The diffusion coefficient as a function of
    amplitude (orange) is enhanced in a specific amplitude region when
    the hollow beam is turned on (from solid to dashed line). A
    corresponding reduction in beam tail population (black) is created
    (from solid to dashed line).}
  \label{fig:active_halo_control}
\end{figure*}

In circular accelerators and storage rings, beam quality can be
affected by the interplay of external excitations with machine
lattice. This work, through calculations and experiments, focuses on
how a certain class of resonant excitations influences beam dynamics
and to which extent these excitations can cause beam losses, emittance
growth, or changes in the particle beam distributions. Besides their
general relevance to the topic of complex nonlinear dynamics, these
studies were motivated by the need to assess the effects of a pulsed
hollow electron lens for active beam halo control.

Considering past, current and future high energy colliders, each new
machine has represented a considerable leap in stored beam energy
(Table~\ref{tab:stored_energy}). Furthermore, recent measurements at
the LHC show that the tails of the transverse beam distribution are
overpopulated compared to a Gaussian distribution. This results in a
considerable amount of energy being stored in the beam tails. In
particular, in the case of the LHC, about 5\% of the beam population
is stored in the tails (i.e., above 3.5$\sigma$, where $\sigma$ is the
standard deviation of the Gaussian beam core), compared to 0.22\% in
an ideal Gaussian distribution, leading to 19~MJ of stored energy for
nominal LHC parameters and 34~MJ in the case of
HL-LHC~\cite{helreview_valentino}. This leads to the conclusion that a
mechanism is needed to deplete the beam tails in a controlled
manner. Further information on the needs for halo control in LHC can
be found in Ref.~\cite{helreview}.

The most direct approach is to decrease the collimator gaps or to
periodically scrape the tails. However, this is not feasible, as it
would generate unacceptably large loss spikes and possibly component
damage. Most promising are methods which increase the diffusion speed
in the region of the halo particles, resulting in a smooth and
continuous removal of the high amplitude tails, while leaving the core
of the beam unperturbed. The diffusing halo particles are then
intercepted by the collimation system and removed. This concept is
also referred to as active halo control, designed to enhance a
conventional passive system, which is still needed to robustly
intercept the halo particles. An illustration of the concept is shown
in Fig.~\ref{fig:active_halo_control}.

In a recent review, the need for such an active halo control system
for HL-LHC has been assessed, with the conclusion that it would
considerably increase the operational margins and reduce the risks for
machine protection~\cite{helreview}. In view of the need of active
halo control for HL-LHC and for future high power accelerators, like
HE-LHC and FCC-hh~\cite{helhcparam2011, fcc_coll_ipac2017}, different
active halo control methods have been
studied~\cite{helreview_bruce}. The hollow electron lens (HEL) is
considered the most established, flexible and suitable technology for
the HL-LHC~\cite{hel_tevatron_stancari, helreview}.

However, the beneficial effects of an HL-LHC HEL for machine
protection and for collimation cannot come at the expense of
performance degradation due to losses or emittance growth in the beam
core. In standard electron-lens operation, a proton bunch sees the
same electron current at every revolution. It is also possible to have
different currents for different groups of bunches. Under these
conditions, the imperfections of the hollow beam have a negligible
effect. On the other hand, in order to extend the range of achievable
removal rates, pulsed operation is also being considered. In this
case, different currents can be set to act on the same bunch at each
turn. If a residual field is present at the location of the beam core,
core particles can be exposed to resonant transverse kicks and to
noise.

In this paper, we concentrate on the experimental and numerical
assessment of possible detrimental effects on the beam core of a
pulsed electron lens. Section~\ref{sec:hel} gives an introduction to
the concept of HELs and summarizes the design parameters of the HL-LHC
HELs. Section~\ref{sec:core} is dedicated to the sources of residual
fields from the HEL in the core region. Sections~\ref{sec:exp}
and~\ref{sec:sim} describe the experimental conditions and the setup
of the numerical tracking simulations. Results and comparisons between
simulations and measurements are given in Section~\ref{sec:simex},
with discussion and summary in Section~\ref{sec:sum}.

\section{Hollow electron lens for HL-LHC}
\label{sec:hel}

\begin{figure*}
  \includegraphics[width=0.8\textwidth]{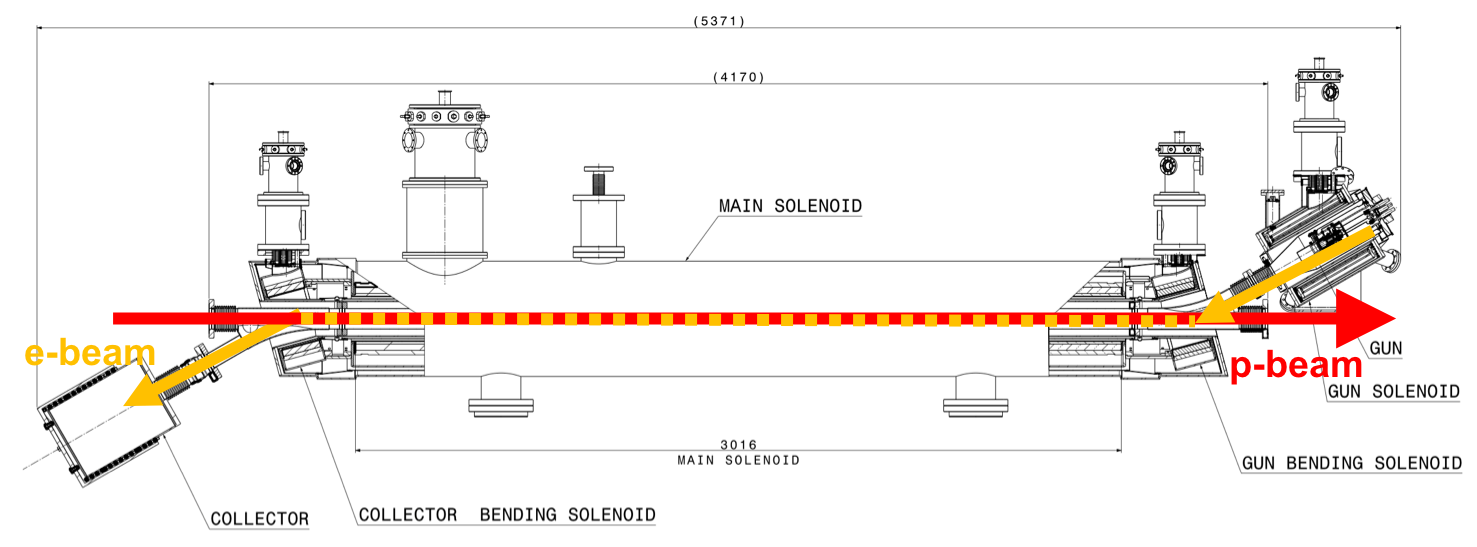}
  \caption{Layout of the hollow electron lens for HL-LHC. (Courtesy of
    CERN EN-MME mechanical engineering group.)}
  \label{fig:hel_layout}
\end{figure*}

\begin{figure}
  \includegraphics[width=0.8\columnwidth]{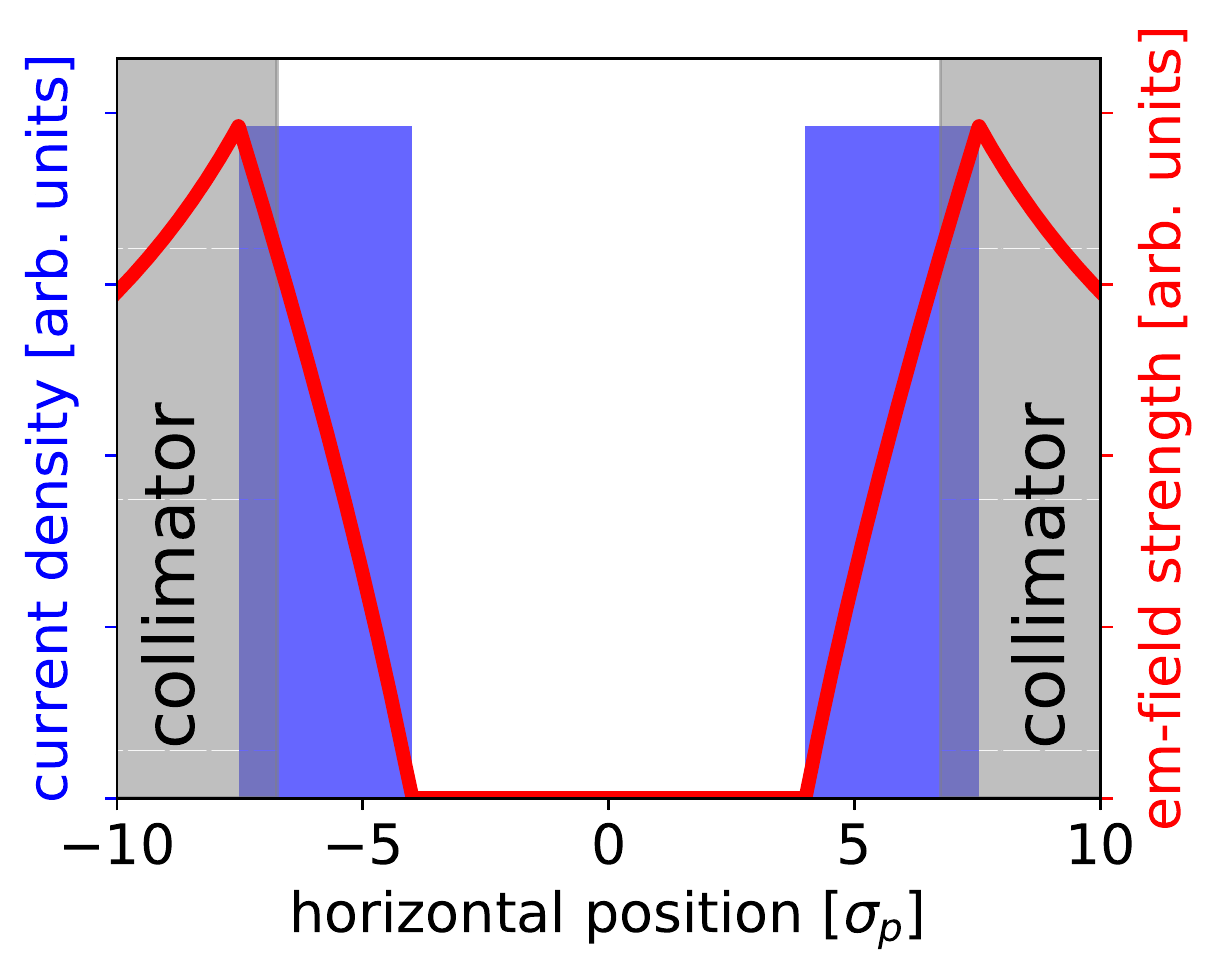}
  \caption{Illustration of the hollow electron beam charge
    distribution (blue), of the magnitude of the transverse kick
    experienced by the proton beam (red), and of the position of the
    primary collimators (gray).}
  \label{fig:hel_field}
\end{figure}

\begin{table}
  \caption{HL-LHC design parameters at top energy~\cite{hlcdr} and
    parameters relevant to the HEL. Optics parameters at the HEL are
    based on a position of $-$40~m for Beam~1 (B1) and $+$40~m for
    Beam~2 (B2) from the interaction point IP4, using HL-LHC optics
    V1.3 with $\beta^{*} = \q{0.15}{m}$~\cite{hlv13}.}
  \label{tab:hllhc_param}
  \begin{ruledtabular}
    \begin{tabular}{lccc}
      Beam parameters & \multicolumn{2}{c}{Value} & Unit\\
                       & B1 & B2 & \\
      \colrule
      Beam energy,  $E_{p}$  &  \multicolumn{2}{c}{7} & TeV\\
      Number of bunches, $n_b$ & \multicolumn{2}{c}{2748} &  \\
      Bunch population, $N_b$ & \multicolumn{2}{c}{$2.2\times 10^{11}$} & \\
      Normalized emittance, $\epsilon_{N,x/y}$ & \multicolumn{2}{c}{2.5} & $\mu$m\\
      Bunch spacing & \multicolumn{2}{c}{25} & ns\\
      \colrule
      \multicolumn{4}{l}{Optics parameters at HEL (B1)\footnote{As the
      Twiss parameters at IP4 do not change during the entire squeeze
      of the optical functions, and IP4 and the HEL are only separated
      by a drift space, the Twiss parameters stay constant also at the
      HEL during the squeeze.}} \\
      \colrule
      $\beta_{x}$ at HEL  & 197.5 & 280.6 & m\\
      $\beta_{y}$ at HEL & 211.9 & 262.6 & m\\
      Dispersion $D_{x}$ at HEL & 0.0& 0.0 & m\\
      Dispersion $D_{y}$ at HEL & 0.0& 0.0 & m\\
      Proton beam size $\sigma_{p,x}$ at HEL & 0.26 & 0.31& mm \\
      Proton beam size $\sigma_{p,y}$ at HEL & 0.27 & 0.30 &mm \\
      \multicolumn{4}{l}{Scale of scraping positions}\\ \hspace{6ex}$\sigma_{p}=\max(\sigma_{p,x},\sigma_{p,y})$ & 0.27& 0.31 & mm\\
    \end{tabular}
  \end{ruledtabular}
\end{table}

\begin{table}
  \caption{HL-LHC hollow electron lens parameters, as defined in
    Ref.~\cite{hel_cdr}.}
  \label{tab:hel_param}
  \begin{ruledtabular}
    \begin{tabular}{lcc}
      Geometry & Value& Unit\\
      \colrule
      Length, $L$    &  3 & m\\
      % 3-8 sigmap with 3.5 um -> 3.55-9.47 for 2.5 um emittance
      Desired range of scraping positions & 3.5--9.5 &$\sigma_p$\\
      \colrule
      Magnetic fields & & \\
      \colrule
      Gun solenoid, $B_g$ & 0.2--0.4 & T\\
      Main solenoid, $B_m$ & 2--6 & T\\
      Collector solenoid, $B_c$ & 0.2--0.4 & T\\
      Compression factor, $k \equiv \sqrt{B_m/B_g}$ & 2.2--5.5 &  \\
      \colrule
      Electron gun & & \\
      \colrule
      Peak yield $I_e$ at 10~keV & 5.0 & A\\
      Gun perveance, $P$ & 5 & $\mathrm{\mu A / V^{3/2}}$\\
      Inner/outer cathode radii, $R_1/R_2$ & 6.75/12.7 & mm\\
      \colrule
      High-voltage modulator & & \\
      \colrule
      Cathode-anode voltage & 10.0 & kV\\
      Rise time (10\%-90\%) & 200 & ns \\
      Repetition rate & 35 & kHz
    \end{tabular}
  \end{ruledtabular}
\end{table}

\subsection{General overview}
\label{sec:hel:intro}

Electron lenses are based upon continuous or pulsed low-energy,
magnetically confined electron beams~\cite{Shiltsev:PRSTAB:2008,
  Fischer:PRL:2015, Gu:PRAB:2017, Shiltsev:elens-book:2016}. The
electron beam is generated in an electron gun, guided and confined by
strong solenoids and finally dumped in a collector. As an example, the
conceptual design of the HL-LHC HEL is shown in
Fig.~\ref{fig:hel_layout}.

The circulating beam (protons in the LHC case) is affected by the
electromagnetic field of the electron beam. For the application of
active halo control, the electron beam needs to generate an
electromagnetic field only at the location of the halo particles. This
field distribution can be achieved, for instance, by using a hollow
charge distribution in radius $r=\sqrt{x^2+y^2}$, uniformly
distributed between inner radius $R_1$ and outer radius $R_2$
(Fig.~\ref{fig:hel_field}). In this case, the circulating proton beam
experiences the following radial kick $\theta(r)$:
\begin{equation}
  \label{eq:field_1}
  \theta(r)=\frac{f(r)}{(r/R_2)}\cdot \theta_{\rm max},
\end{equation}
where $f(r)$ is a shape function with
\begin{equation}
  \label{eq:field_2}
  f(r) =
  \begin{cases}
    0 &,\quad r< R_1,\\
    \frac{r^2-R_1^2}{R_2^2-R_1^2} &,\quad R_1 \leq r < R_2,\\
    1 &,\quad R_2 \leq r
  \end{cases}
\end{equation}
and $\theta_{\rm max}=\theta(R_2)$ is the maximum kick angle given by
\begin{equation}
  \label{hel_kick_max}
  \theta_{\rm max} = \theta(R_2) =
  \frac{2LI_T(1\pm\beta_e\beta_p)}{4\pi\epsilon_0  \cdot
    \left(p_0/q\right)_p \cdot \beta_e\beta_p
    c^2}\cdot\frac{1}{R_2},
\end{equation}
with $L$ the length of the HEL, $I_T$ the total electron beam current,
$\beta_{e}$ and $\beta_{p}$ the relativistic velocity parameters of
electrons and protons, $\left(p_0/q\right)_p = \left(B\rho\right)_p$
the magnetic rigidity for the proton beam reference particle, $c$ the
speed of light and $\epsilon_0$ the vacuum permittivity. The
$\pm$-sign in Eq.~\ref{hel_kick_max} represents the two cases of the
electron beam traveling in the direction of the proton beam
($v_e v_p>0$) leading to ``$-$" or in the opposite direction
($v_e v_p<0$) leading to ``$+$". For hollow electron beam collimation,
electrons and protons are chosen to counterrotate, so that the
magnetic and electric kicks add up. (For simplicity, the dependence of
the electron axial velocity on radius is neglected in
Eq.~\ref{hel_kick_max}.)

In the case of HL-LHC HEL design parameters
(Table~\ref{tab:hllhc_param} and Table~\ref{tab:hel_param}), the
maximum kick is:
\begin{equation}
  \label{eqn:helkick}
  \theta_{\rm max,B1} = 392~\rm{nrad}
\end{equation}
for an inner radius of $R_1 = 4\sigma_p$, outer radius $R_2 =
7.5\sigma_p$, peak current of $I_e = \q{5.0}{A}$, using the Beam~1
lattice of the LHC. Similar values are obtained for Beam~2.

\subsection{Operation modes and effects on the beam core}
\label{sec:hel:core}

For the HEL, two modes of operation are currently under consideration:
the continuous mode (also referred to as `DC', or direct current, in
this paper) as standard operation mode, described above; and the
pulsed mode. The main benefit of pulsed HEL operation is the increase
in halo removal rates. A wider range of removal rates may become
important under operating conditions with small nonlinearities, in
particular low chromaticity and octupole current, when the DC mode may
be too slow~\cite{hel_halo_hllhc_fitterer, hl_halo_ipac2017}.

Two different pulsing patterns are considered for the HL-LHC. In both
cases, at each passage, a given bunch sees a different electron-lens
current and, therefore, experiences a different transverse kick. The
patterns are defined as follows:

\begin{itemize}

\item \textbf{random excitation:} The extraction voltage in the
  electron gun is modulated according to the
  following expression:
  \begin{equation}
    \label{eq:random}
    U_{\mathrm{e-gun}} = (1-a) \cdot U_{\mathrm{max}} +
    a \cdot \eta \cdot U_{\mathrm{max}},
  \end{equation}
  where $U_{\mathrm{max}}$ is the maximum voltage, $a$ is the
  modulation strength, with $a \in [0,1]$, and $\eta$ is a uniformly
  distributed random number in the interval~$[0,1]$. Simulations and
  experiments, discussed below, were usually conducted with $a = 1$.

\item \textbf{resonant excitation:} The electron beam is switched on
  only every \kth\ turn. The excitation can be represented
  by the following expression:
  \begin{eqnarray}
    \label{intro:eqn:1}
    f(t) & = & \sum_{n=-\infty}^{+\infty}\delta\left[t-n\cdot(kT)\right],
  \end{eqnarray}
  where $n$ is the turn number and $T$ is the revolution period. Its
  Fourier representation is
  \begin{equation}
    \label{intro:eqn:2}
    f(t) = \Sh_{kT}(t) = \frac{1}{kT}\sum_{n=-\infty}^{+\infty}e^{2\pi i f_nt},
  \end{equation}
  with $f_n = n \cdot f_{\rm rev} / k$ and where $\Sh_{kT}$ is the
  Dirac comb. In general, \kthtp\ drives \kth-order
  resonances~\cite{md_sim_hel_res_ex_fitterer}. This type of pulsing
  pattern was used in the Tevatron during regular collider operations
  for abort-gap cleaning~\cite{hel_tevatron_abortgap_zhang}.
\end{itemize}

For an axially symmetric electron lens, the field at the beam core
vanishes. Effects on the proton core arise from imperfections, with
two main sources: the injection and extraction bends of the HEL
(discussed below in Section~\ref{core:sec:1}), where the electron beam
crosses the proton beam; and distortions of the electron beam profile
during its propagation under magnetic focusing and space charge
(Section~\ref{core:sec:2}). Both sources result in nonlinear
kicks~\cite{hel_bends_stancari, hel_model_polynomial_morozov}. In
continuous operation, these nonlinear kicks are usually much smaller
than the machine nonlinearities. Tolerances on imperfections are
therefore not particularly stringent. The picture changes
significantly in case of pulsed operation. If the electromagnetic
field does not vanish at the proton beam core, noise or resonant kicks
are transferred not only to the halo particles, as intended, but also
to the beam core. Tolerances on the residual fields in this case
become much more stringent. Studies of the effects of the HEL on the
beam core therefore focus on this mode of operation, which is also the
main subject of this paper.

\section{Sources and estimates of residual fields}
\label{sec:core}

\begin{figure*}
  \begin{tabular}{cc}
    \includegraphics[width=0.48\textwidth]{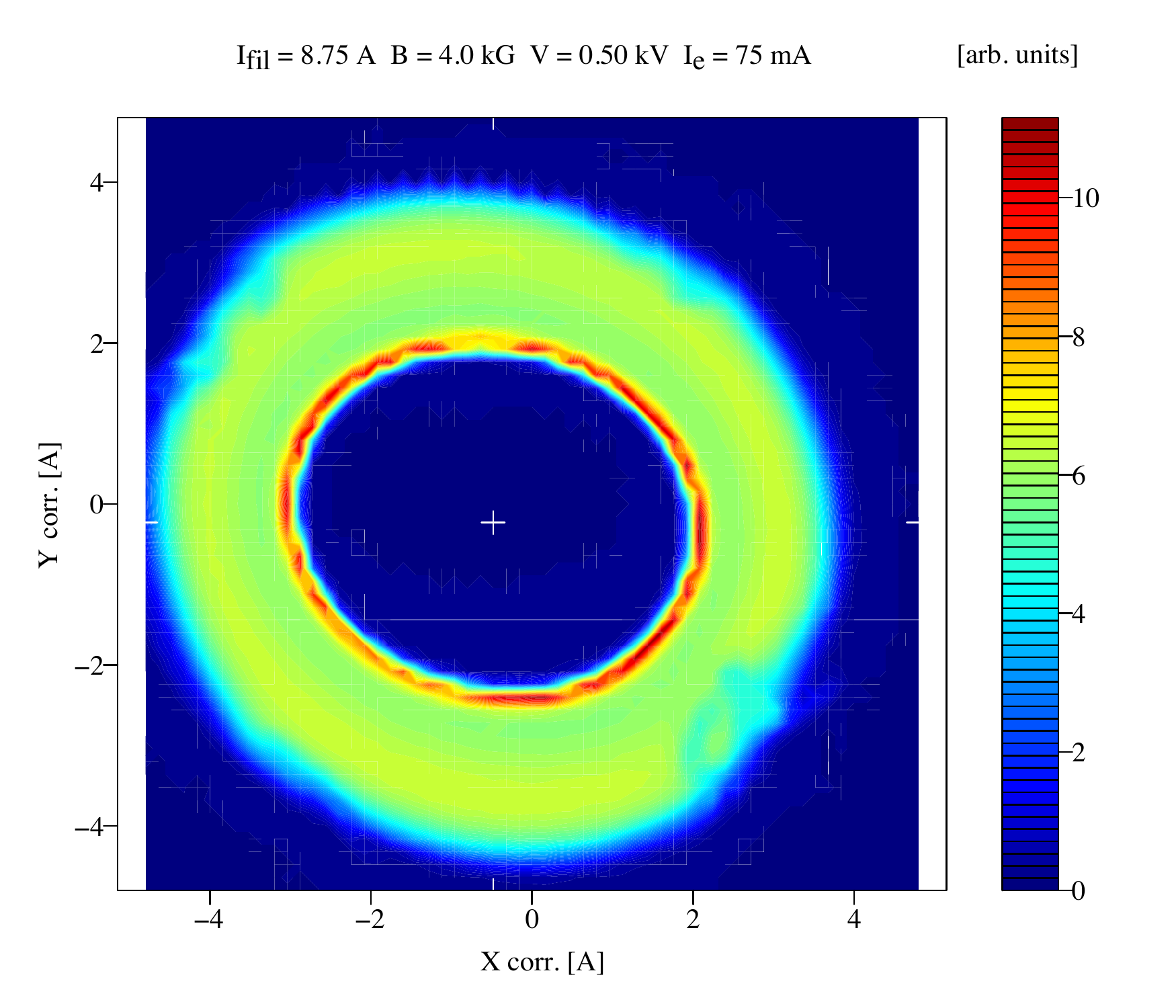} &
    \includegraphics[width=0.4\textwidth]{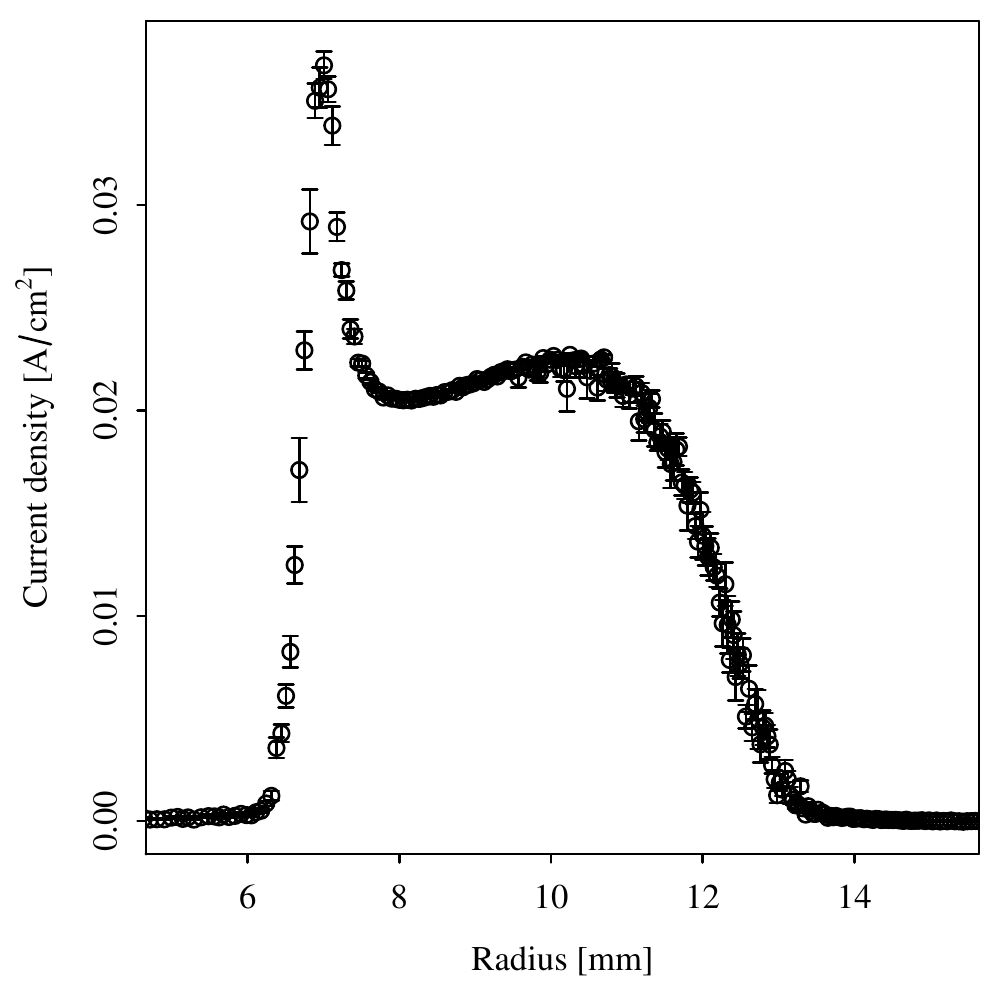}\\
  \end{tabular}
  \caption{Example of current-density distribution measurements for
    the hollow electron gun prototype CHG1b, taken at the Fermilab
    electron lens test stand in
    2017~\cite{hel_res_field_stancari_2017}: 2-dimensional transverse
    profile measurement (left) and calculated 1-dimensional radial
    projection (right).}
  \label{core:fig:0}
\end{figure*}

\begin{figure*}
  \begin{tabular}{cc}
    \includegraphics[width=0.4\textwidth]{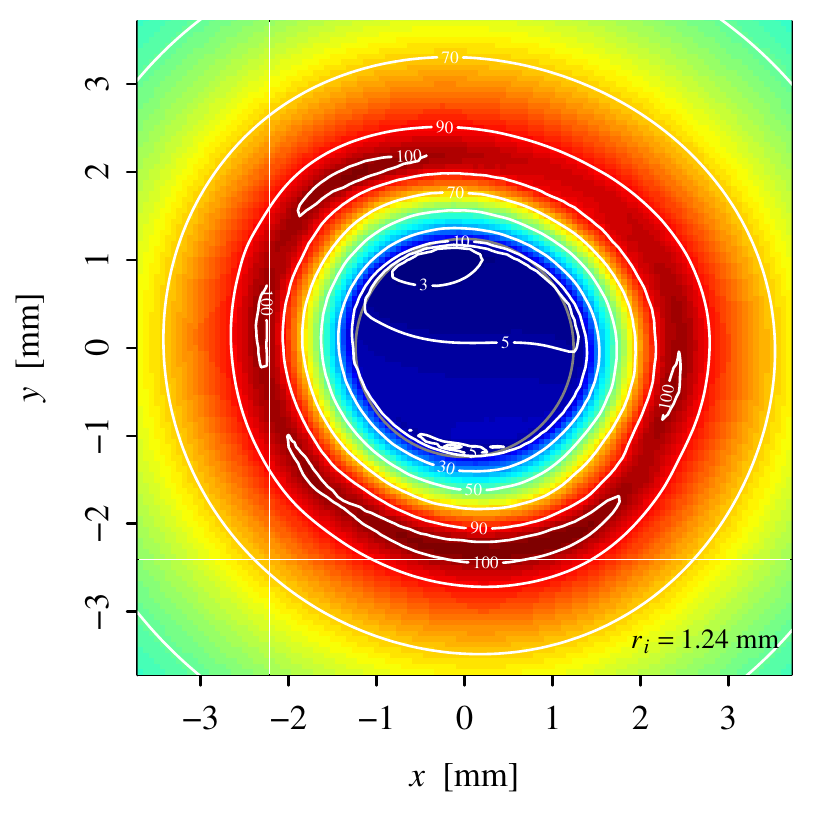} &
    \includegraphics[width=0.4\textwidth]{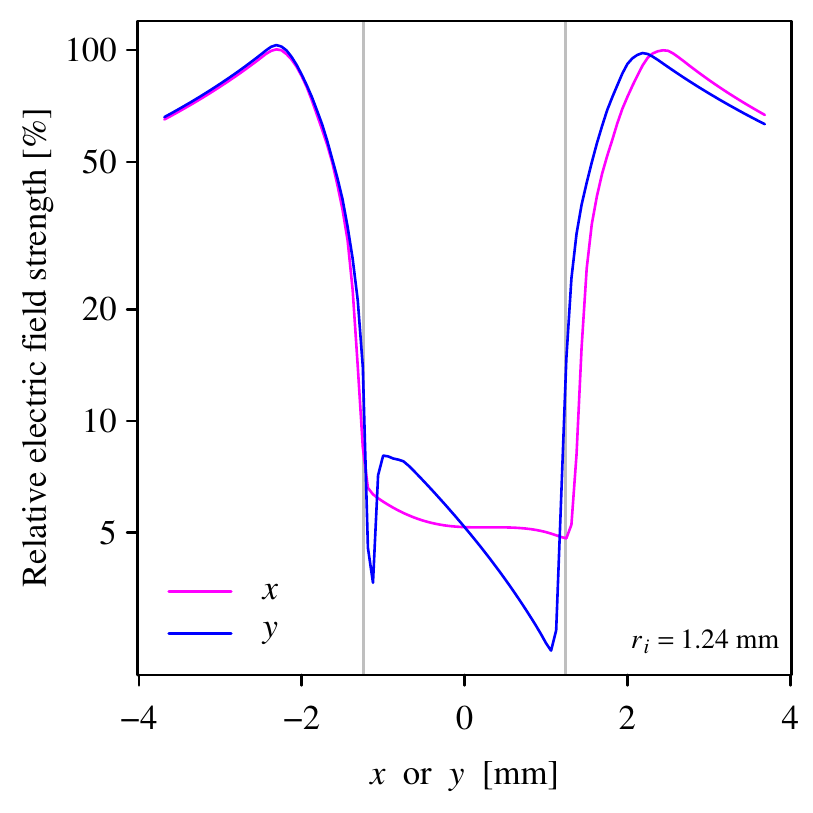}\\
  \end{tabular}
  \caption{Calculated relative electric field for the hollow electron
    gun CHG1b in the transverse $x$-$y$ plane (left) and as
    1-dimensional cuts through the $x$ and $y$ axes (right). The field
    calculations are based on measurements at the Fermilab
    electron-lens test stand combined with \code{warp} calculations of
    the electric potentials and fields in a cylindrical beam pipe.}
  \label{core:fig:1}
\end{figure*}

Parasitic kicks on the proton beam core can be due to electron-beam
profile imperfections in the overlap region, or to the injection and
extraction toroidal bends, where electrons and protons overlap, as
shown in the layout of Fig.~\ref{fig:hel_layout}.

Because no HEL is currently installed in the LHC, the kicks on the
proton beam core must be emulated by other devices to determine their
effects experimentally in a given machine and to guide design and
tolerances. In particular, during the experiments presented in this
paper, the LHC transverse damper system (described in detail in
Section~\ref{sec:adt}) was used for this purpose. This system can
generate transverse dipole kicks with a wide range of excitation
patterns.

For comparison with experiments and simulations, here we estimate to
first order the magnitude of the dipole kicks that may be expected
from the HEL. As one can see below, the contribution from the central
region is in general dominating.

\subsection{Kicks from injection and extraction bends}
\label{core:sec:1}

To estimate the dipole component of the kicks that originate from the
injection and extraction bends, we used the approach described in
Ref.~\cite{hel_bends_stancari}. The bends are modeled as a bent
cylindrical pipe with a static charge distribution of electrons. The
resulting electric field is calculated, with the vacuum pipe as
boundary, using the solvers of the \code{warp} particle-in-cell
code~\cite{warp, Friedman:IEEE-plasma:2014}. (The contribution of the
magnetic field generated by the electron current was neglected in this
study.) The field, integrated over the trajectory of the protons, is
then translated into a symplectic kick map.

In case of a U-shaped electron lens, where electron gun and collector
are on the same side, the transverse (and possibly pulsed) dipole
kicks generated by the electron charge at the entrance and exit add
up. For an S-shaped electron lens, on the other hand, where gun and
collector are on opposite sides, these kicks compensate each
other. For this reason, an S-shape was chosen for the HL-LHC HEL
(Fig.~\ref{fig:hel_layout}). A disadvantage of the S-shape is that the
static magnetic kicks due to the toroidal sections do add up, but they
can be compensated by conventional dipole correctors, especially in a
high-energy machine.

In the case under study of an S-shaped electron lens, residual
uncompensated kicks arise from differences in electron charge
distribution between the injection and extraction bends. Here we
conservatively assume 10\% fluctuations between the entrance and exit
and, furthermore, that these differences add up.

The maximum values of the integrated electric fields calculated in
Ref.~\cite{hel_bends_stancari}, based upon an electron beam of 1~A at
5~keV, are
\begin{equation}
  \int_{z_1}^{z_2} E_{x,y} \, dz= \q{10}{kV}.
\end{equation}
Scaling to HL-LHC and HEL design parameters
(Table~\ref{tab:hllhc_param} and Table~\ref{tab:hel_param}) yields the
following integrated field and corresponding kick:
\begin{equation}
  \int_{z_1}^{z_2} E_{y} \, dz = \q{36}{kV} \Rightarrow \Delta y' = \q{5.1}{nrad},
\end{equation}
as described in detail in
Ref.~\cite{md_sim_hel_res_ex_fitterer}. Assuming a residual difference
of 10\% between entrance and exit, the expected kick is approximately
\begin{equation}
  \label{eqn:kick_bends}
  \Delta x', \Delta y' = \q{0.5}{nrad}.
\end{equation}

\subsection{Kicks in the central overlap region}
\label{core:sec:2}

For a perfectly annular and axially symmetric electron beam profile,
the electromagnetic field in the region of the proton beam core
vanishes. This is expressed by Eq.~\ref{eq:field_2} and is illustrated
in Fig.~\ref{fig:hel_field}. Fields at the location of the proton beam
core can arise if the axial symmetry is broken.

Recently, a hollow electron gun prototype for the LHC (called CHG1b)
was characterized at the Fermilab electron lens test
stand~\cite{hel_test_stand_fnal}. An example of a measurement of the
electron beam current density is shown in Fig.~\ref{core:fig:0}.

In the test stand, only resistive solenoids are available. One can
estimate the fields generated by the compressed electron beam profile
in the superconducting solenoids of the HL-LHC HEL using a combination
of experimental measurements and calculations.

Experimentally, it was verified that the current-density profiles
scale with electron beam current and confining axial field according
to space-charge evolution~\cite{Jo:PoP:2018,
  hel_res_field_stancari_2017}. Specifically, the same profile is
obtained for a given family of experimental conditions with constant
ratio $\sqrt{V} / B$, where $V$ is the accelerating voltage and $B$ is
the axial field. This ratio is proportional to the space-charge
evolution number $g = \omega_D \cdot \tau \propto \sqrt{V}/B$,
representing the number of $\vec{E} \times \vec{B}$ rotations in the
propagation time~$\tau$, with
$\omega_D \equiv \omega_p^2 / (2 \omega_c)$ the diocotron frequency,
$\omega_p$ the plasma frequency, and $\omega_c$ the cyclotron
frequency of the magnetically confined electrons in the
solenoid~\cite{Davidson:nonneutral-plasmas:2001}.

For a given HL-LHC HEL configuration, the corresponding
current-density profile measured in the Fermilab test stand is used as
input to calculate the electromagnetic fields. For the purpose of
estimating the residual fields, the measured distribution is
compressed to the inner electron beam radius of 1.24~mm, corresponding
to $4\sigma_p$, as described in Table~\ref{tab:hllhc_param} and
Table~\ref{tab:hel_param}. A distribution with about $65\,000$
particles is generated according to the measured current-density
profile. As boundary condition, the LHC inner beam pipe radius
$b = \q{30}{mm}$ is used. The potential and fields are then calculated
with \code{warp}~\cite{warp, Friedman:IEEE-plasma:2014}. The resulting
relative electric field strengths are shown in
Fig.~\ref{core:fig:1}. The electric field is obviously proportional to
the charge density of electrons. Its relative strength is rather
insensitive to the hollow beam radii, as long as these remain small
compared to the beam pipe radius. Further details on the measurements
and \code{warp} simulations can be found in
Ref.~\cite{hel_res_field_stancari_2017}.

Using these measurements and calculations of the relative electric
field strengths in the transverse plane (Fig.~\ref{core:fig:1}), we
obtain a median field in the hole of
\begin{equation}
  \langle E_\mathrm{hole} \rangle / E_\mathrm{max} =  5\%.
\end{equation}
For a maximum kick of 392~nrad, as derived in Eq.~\ref{eqn:helkick},
the estimated dipole kick amplitude at the proton beam core is therefore
approximately
\begin{equation}
  \label{eqn:kick_central}
  \Delta x' , \Delta y' = \q{20}{nrad}.
\end{equation}
The relative magnitude of the residual field depends on several
factors, including cathode quality, electron gun geometry, solenoid
field configuration, space-charge evolution, etc. It can be improved,
if needed.

For the purposes of this paper, only the approximate magnitude of the
residual dipole kick is considered. In general, the field map,
including its multipolar components, can be parameterized in
symplectic form, as described for instance in
Ref.~\cite{hel_bends_stancari}, and used in tracking codes to estimate
the effects on the circulating beam.

\section{Experimental setup}
\label{sec:exp}

\begin{table*}
  \caption{Beam parameters and machine configuration for the two
    resonant excitation experiments of~2016
    and~2017~\cite{resexmd2016, resexmd2017}. The plane of the
    excitation is abbreviated as H for horizontal, V for vertical and
    H+V for horizontal and vertical at the same time. MOF and MOD
    refer to the focusing and defocusing Landau octupoles.}
  \label{tab:md_param}
  \begin{ruledtabular}
    \begin{tabular}{lcc}
      Parameter & Experiment 2016 & Experiment 2017  \\
      \colrule
      beam &\multicolumn{2}{c}{Beam~1} \\
      beam energy &\multicolumn{2}{c}{injection energy, 450 GeV} \\\hline
      single bunch intensity &\multicolumn{2}{c}{$0.7\times10^{11}$} \\
      normalized emittance &\multicolumn{2}{c}{$2.5$--\q{3.5}{\mu m}} \\
      $4\sigma$ bunch length & \multicolumn{2}{c}{1.3~ns}\\
      $1\sigma$ bunch length & \multicolumn{2}{c}{9.7~cm}\\
      number of bunches & $12 \times 4 = 48$ & $3 \times 72 = 216$ \\
                &  & (+ 1 pilot + 12 nominal) \\\hline
      injection optics, $\beta^* = \q{11}{m}$ & standard optics 2016 & standard optics 2017\\
      Landau-damping octupoles  & \multicolumn{2}{c}{$I_{\mathrm{MO}}
                                  = \q{+19.6}{A}$ for MOF circuit and}\\
                & \multicolumn{2}{c}{$I_{\mathrm{MO}} = \q{-19.6}{A}$
                  for MOD circuit (standard 2016 settings)} \\ \hline
      working point $(Q_x, Q_y)$ & (64.28, 59.31) & (62.27, 60.295) \\
      chromaticity $(Q'_x, Q'_y)$ & \multicolumn{2}{c}{(+15, +15)}\\ \hline
      pulsing patterns  & 7th~turn H & 7th~turn H, V, H+V \\
                & 10th~turn V & 8th~turn H, V, H+V \\
                & &  random  H, V, H+V\\
    \end{tabular}
  \end{ruledtabular}
\end{table*}

\begin{figure*}
  \begin{tabular}{c}
    \multicolumn{1}{l}{\emph{2016 experiment}} \\
    \includegraphics[width=0.9\textwidth]{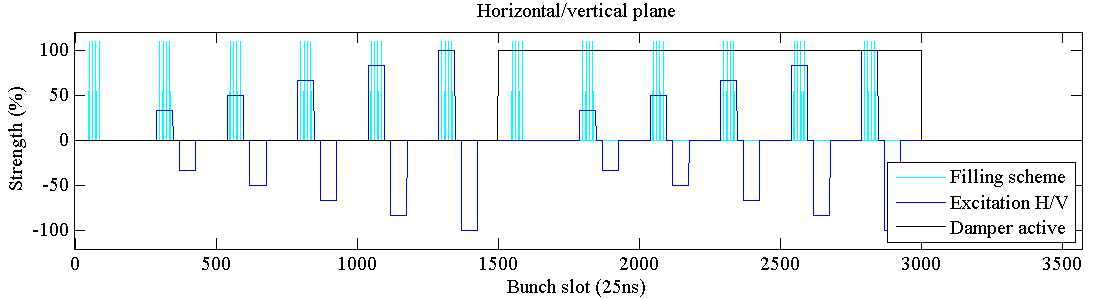} \\
    \\
    \multicolumn{1}{l}{\emph{2017 experiment}} \\
    \includegraphics[width=0.9\textwidth]{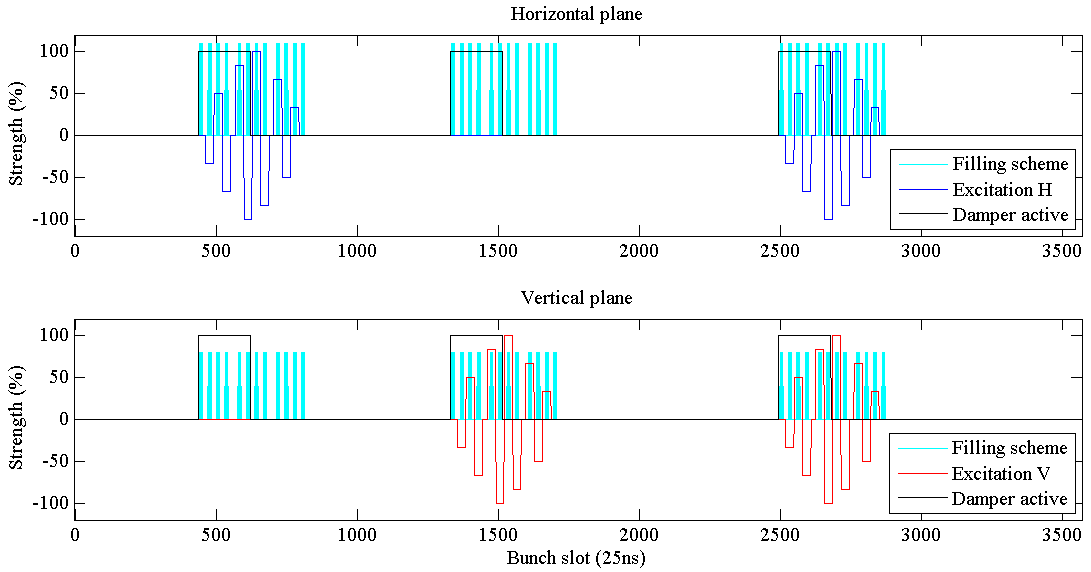}
  \end{tabular}
  \caption{Bunch filling scheme and excitation patterns for the
    2016~(top) and 2017~(bottom) LHC experiments. In 2016, a total of
    48~bunches was used, whereas in 2017 there were 216~bunches. Each
    bunch is represented by a vertical cyan bar. The bunches were
    grouped in subsets of~4 in 2016 and in subsets of~6 in 2017. Each
    subset experienced the same excitation pattern and amplitude. The
    excitation amplitudes and relative phases are shown in blue or
    red. In 2016, the excitation was only applied in one plane. In
    2017, more injected bunches were allowed without compromising
    machine protection, so it was possible to test all~3 excitation
    planes in the same fill. The transverse damper was active on half
    of the bunches, indicated by the black lines.}
  \label{fig:fill}
\end{figure*}

\begin{figure*}
  \includegraphics[width=0.9\textwidth]{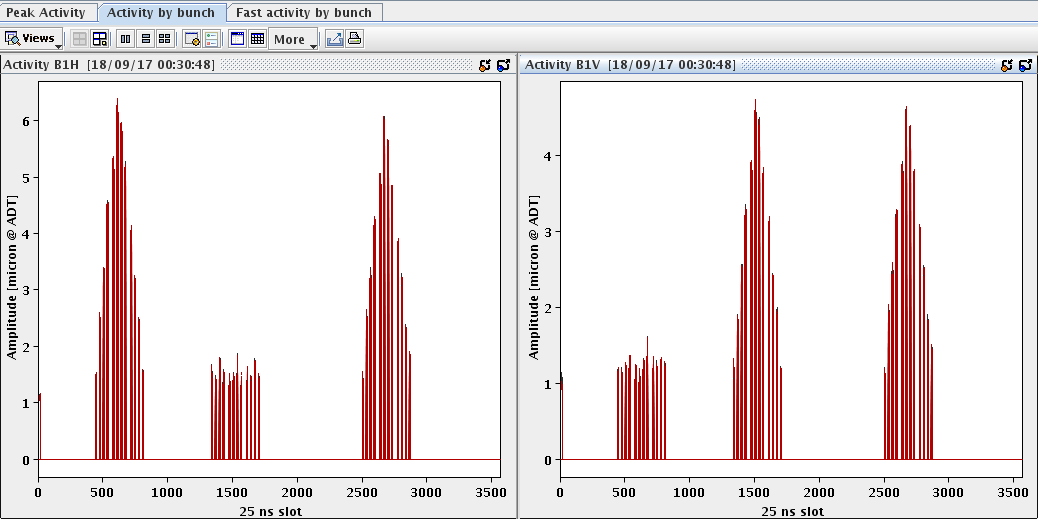}
  \caption{Example of bunch-by-bunch beam centroid motion for Beam~1
    in the horizontal (B1H, left) and vertical (B1V, right) planes,
    detected by the real-time transverse activity monitor during the
    2017 experiment, when the beam was excited every 8th~turn.}
  \label{fig:fill_meas}
\end{figure*}

\subsection{Overview}
\label{sec:exp_sum}

The purpose of the experiments at the LHC is to quantify the effects
of a pulsed excitation on the proton beam core. In addition, these
measurements provide an experimental basis to guide the design and
tolerances on the residual HEL fields at the location of the beam
core, in case resonant excitation is needed for HL-LHC.

Two experiments were conducted, one in 2016~\cite{resexmd2016} and one
in 2017~\cite{resexmd2017}. Beam and machine parameters are summarized
in Table~\ref{tab:md_param}.

During the experiments, losses were measured with the fast beam
current transformers (FBCTs). Transverse beam profiles and emittances
were provided by the beam synchrotron radiation telescope
(BSRT)~\cite{Trad:PhD:2015}. All instruments were capable of
delivering bunch-by-bunch data. The data analysis of the BSRT profiles
is quite involved. In this paper, we focus on the direct comparison of
the resulting profiles. A detailed description of the profile analysis
can be found in Ref.~\cite{bsrtprofinj}, with individual experiments
reported in Refs.~\cite{resexmd2016, resexmd2017}.

The choice of excitation patterns for the experiments was guided by
losses and emittance growths calculated in numerical tracking
simulations, described below. It was chosen to study experimentally
the two pulsing patterns with the largest calculated effects on the
beam (7th- and \tenthtp), one pattern with no effect (\eighthtp), and
the random excitation. In order to quantify the reproducibility of the
results under different machine configurations, one pulsing pattern
(\seventhtp) was tried first in~2016 and then repeated in~2017.

\subsection{Excitations with the transverse damper and bunch filling schemes}
\label{sec:adt}

The primary function of the LHC transverse damping and feedback
system, also known as ADT, is to mitigate injection oscillations and
to actively damp the coupled-bunch instabilities driven by machine
impedance~\cite{adt_sum_2008, adt_sum_2011}. The main building blocks
of the system are the following: strip-line pickups at positions Q7
and Q9 near Interaction Point~4 (IP4) of the LHC, which are connected
to the beam position measurement modules at the surface; the digital
signal processing modules (mDSPU); and a set of tetrode power
amplifiers feeding electrostatic kickers in the same radio-frequency
sector of the LHC (IP4).

Because of its flexibility and state-of-the-art hardware, the system
is being routinely used for sophisticated beam excitations. These
include abort- and injection-gap cleaning, excitation of individual
bunches for tune and linear-coupling measurements, and other special
modes of operation for dedicated experiments in the LHC.

The transverse feedback is in general active during all phases of LHC
operation. The typical machine cycle requires a short damping time of
10--20 turns (high feedback gain) for injection oscillation
damping. During the acceleration ramp and during collisions, the
damping time is increased to 50--100 turns (lower feedback gain).

Because the damper is always active in operations, it is critical that
the noise introduced by the system does not cause any measurable
emittance growth, and this fact was verified
experimentally~\cite{adt_noise_emit_2017}. All excitation signals
described in this paper were digitally synthesized in the ADT's
digital signal processing units and are therefore assumed to be
`noise-free.' The observed effects on the beam are attributed to the
applied excitations and the effects of unwanted residual noise are
assumed to be negligible.

The resonant excitation experiment in~2017 involved simultaneous
measurements on 3~groups of 72~bunches, with dedicated excitation
patterns and transverse feedback configurations (i.e., damper active
or damper off) on each subset of 6~bunches. In~2016, a similar
configuration was used, with 48~bunches in total and subsets of
4~bunches. Both schemes are illustrated in Fig.~\ref{fig:fill}. A
total of 5 different amplitudes could be applied simultaneously to
different subsets of bunches, denoted as $n \, \Delta A$, with
$n = 1, \ldots, 5$. In addition, there were reference bunches without
excitation for each transverse feedback configuration and excitation
plane. Observables (losses, emittances, etc.) were averaged over each
subset of bunches experiencing the same excitation and damping
conditions.

In~2017, the excitations in the horizontal and vertical planes were
generated by a different set of signal-processing devices, but
synchronized at a turn-by-turn level. Therefore, the bunches of the
third group of 72 bunches (see Fig.~\ref{fig:fill}) were affected by
the kicks in both horizontal and vertical planes during the same turn.

The experiments were usually split in time into 3 different periods:
\begin{description}
\item[Period 1] No excitation was applied, to allow the beam to reach
  an equilibrium state after injection.
\item[Period 2] The excitation was applied with a first maximum
  excitation amplitude of $A_{\mathrm{max,1}} = 5 \, \Delta A_1$.
\item[Period 3] The maximum excitation amplitude was further increased
  to $A_{\mathrm{max,2}} = 5 \, \Delta A_2 > A_{\mathrm{max,1}}$.
\end{description}
Each period lasted approximately 10--15~minutes, which was considered
long enough to allow the beam to reach its new equilibrium state. In
the discussion of the experimental results (Section~\ref{sec:simex}
below), the 3~periods are labeled according to the maximum excitation
amplitude~$A_\mathrm{max}$, and subsets of bunches with the same
excitation amplitude $n \, \Delta A$ ($n = 1, \ldots, 5$) are grouped
by color.

The proton deflection angle generated by the ADT kicker was calculated
from the kicker geometry and from the excitation voltage on the
deflection plates. The voltage depends on a complex chain of hardware
(digital signal processor, transmission lines, low- and high-power
amplifiers, etc.). In~2016, the kick was estimated from the
operational system parameters and it was assigned an uncertainty of
50\%. In~2017, the excitation voltage could be indirectly measured
once, using the probes mounted on the kickers, with an estimated
uncertainty of 10--15\%. A precise in-situ calibration was not
possible due to the limited machine availability for dedicated
studies.

The maximum ADT kick strength that could be obtained at injection
without risking saturation was approximately 100~nrad. For the
experiments, a maximum nominal kick strength of 96~nrad was chosen.

Figure~\ref{fig:fill_meas} shows an example of the oscillation
amplitude of each bunch centroid in both horizontal and vertical
planes, captured by the LHC real-time transverse activity monitor
during the 2017 experiments, when the beam was excited every
8th~turn. The measured activity shows the typical amplitude of bunch
centroid excursions and reflects the expected excitation pattern.

\section{Numerical tracking simulations}
\label{sec:sim}

\begin{table*}
  \caption{Summary of parameters used in numerical simulations of
    distribution tracking and frequency-map analysis (FMA). Further
    details can be found in Refs.~\cite{md_sim_hel_res_ex_fitterer,
      resexmd2017}.}
  \label{tab:sim_param}
  \begin{ruledtabular}
    \begin{tabular}{lcc}
      Parameter & distribution & FMA \\
      \colrule
      beam &\multicolumn{2}{c}{Beam~1} \\
      beam energy &\multicolumn{2}{c}{450 GeV} \\
      normalized emittance & \q{3.5}{\mu m} & \q{2.5}{\mu m} \\
      $4\sigma$ bunch length & \multicolumn{2}{c}{1.3~ns} \\
      $1\sigma$ bunch length & \multicolumn{2}{c}{9.7~cm} \\
      particle distribution & \parbox{0.4\textwidth}{6D Gaussian
                              distribution with $10^4$ particles}
                               & \parbox{0.4\textwidth}{equally spaced
                                 grid in $x, y$ up to $10\sigma$, with
                                 $(\Delta p / p_0) = 0$} \\
      turns tracked & $10^6$ & $10^4$ \\
      \hline
      optics & \multicolumn{2}{c}{2016 or 2017 injection optics, with
               $\beta^* = \q{11}{m}$ at IP1 and IP5} \\
      machine imperfections & standard errors, with $a_1 = b_1 =
                              0$\footnote{Orbit
                              errors are disabled due to
                              different implementation of the
                              $a_1, b_1$ coefficients in \code{lifetrac} and
                              \code{mad-x}. $b_2$ errors are adjusted to
                              yield an average
                              peak $\beta$-beat of 15\% over 60~seeds, as
                              observed in optics measurements
                              in the LHC.} & no errors \\
      octupoles  & \multicolumn{2}{c}{$I_{\mathrm{MOF}} =
                   \q{+19.6}{A}$, $I_{\mathrm{MOD}} = \q{-19.6}{A}$} \\
      tunes $(Q_x, Q_y)$ & \multicolumn{2}{c}{(64.28, 59.31) for 2016,
                           (62.27, 60.295) for 2017} \\
      chromaticities $(Q'_x, Q'_y)$ & \multicolumn{2}{c}{(+15, +15)} \\
      \hline
      transverse aperture & $5.7\sigma$ &  \\
      longitudinal aperture & $10\sigma$ &  \\
    \end{tabular}
  \end{ruledtabular}
\end{table*}

For the preparation and interpretation of the experiments, two
different types of simulations were performed:
\begin{itemize}
\item tracking of a Gaussian distribution, referred to as
  `distribution tracking' in this paper, to obtain particle loss rates
  and emittance evolution;
\item frequency-map analysis (FMA), to visualize the location and
  intensity of resonances~\cite{fmalaskar}.
\end{itemize}
For both simulation types, the tracking code
\code{lifetrac}~\cite{lifetrac} was used. The simulation parameters
are summarized in Table~\ref{tab:sim_param}. Further details are given
in Refs.~\cite{md_sim_hel_res_ex_fitterer, resexmd2017}.

Quantitative predictions of loss rates and emittance growth are in
general challenging, as both observables are influenced by several
factors. For instance, the natural noise present in the machine
(originating from mechanical vibrations, current ripple in the magnet
power supplies, etc.) may have complex interactions with the external
excitations. In the LHC, noise sources at top energy are well
characterized, whereas they are not well known at injection,
presenting an influential but undefined input for simulations.  An
estimate is obtained by scaling the value at
6.5~TeV~\cite{md1433_noise_top_energy, md_noise_bbLHC} with proton
magnetic rigidity to the injection energy of 450~GeV. This yields a
maximum kick amplitude at the transverse damper of approximately
\begin{equation}
  \label{eq:noise-kick}
  \theta_{\mathrm{random,ADT,max}}(\mathrm{450~GeV}) = \q{6}{nrad}.
\end{equation}

Collective effects, such as intra-beam scattering and electron cloud,
influence the time evolution of losses and emittance as well. To
minimize their effects, experiments were done at low bunch intensities
($0.7 \times 10^{11}$ protons per bunch). The presence of collective
effects was neglected in these simulations.

The results of tracking simulations are discussed below in
Section~\ref{sec:simex} together with the experimental results, to
enhance the understanding and interpretation of the models and
observations.

\section{Results}
\label{sec:simex}

In this Section, we present the experimental observations and compare
them with the results of numerical simulations.  In
Section~\ref{sec:pattern}, we show which pulsing patterns are
predicted to be the most efficient in the LHC.  Next, we present
results for each of the specific excitation patterns that could be
tested experimentally, namely pulsing every 10th~turn
(Section~\ref{sec:simex10}), 7th~turn (Section~\ref{sec:simex7}),
8th~turn (Section~\ref{sec:simex8}), and the random excitation
(Section~\ref{sec:simexran}).  In Section~\ref{sec:damp}, we describe
how the transverse damping system influenced the effects of the
external excitation sources on losses, beam distributions, and
emittances.

\subsection{Dependence on the pulsing pattern}
\label{sec:pattern}

\begin{figure*}
\begin{tabular}{cc}
\includegraphics[width=\halfwidth]{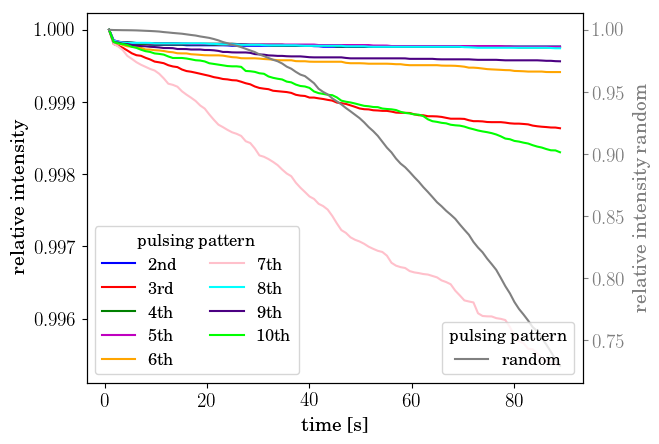}
  &
\includegraphics[width=\halfwidth]{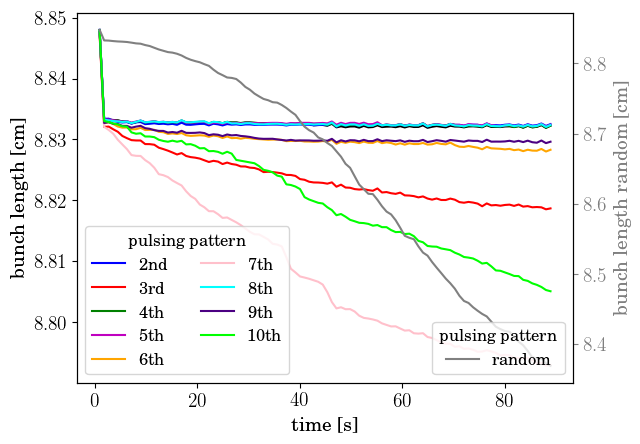}
  \\
\includegraphics[width=\halfwidth]{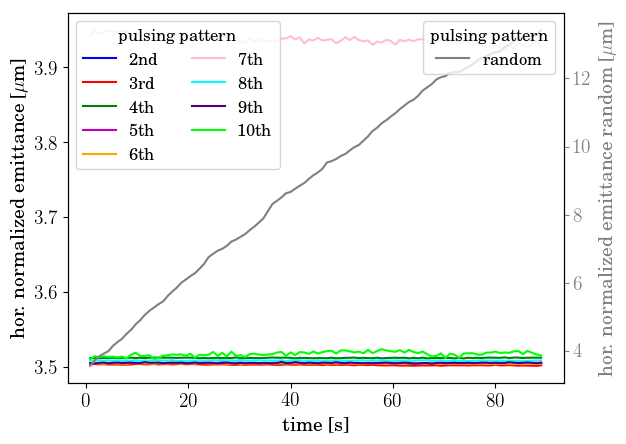} &
\includegraphics[width=\halfwidth]{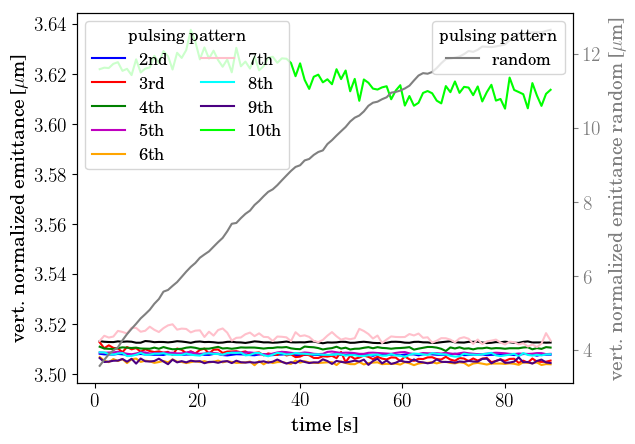}
  \\
\end{tabular}
\caption{Relative intensity (top left), bunch length (top right) and
  horizontal (bottom left) and vertical (bottom right) emittances for
  different pulsing patterns, calculated by distribution tracking
  based on the 2016 injection optics with
  $(Q_x, Q_y) = (64.28, 59.31)$ and standard lattice errors. The
  resonant and random excitations are applied in both planes, with an
  amplitude of 96~nrad. No random noise component is added. Because of
  its much larger effects, the random excitation is shown with
  separate vertical axes.}
\label{fig:patternsim}
\end{figure*}

\begin{figure*}
  \begin{tabular}{cc}
    No excitation & 10th, H+V \\
    \includegraphics[width=\fmawidth]{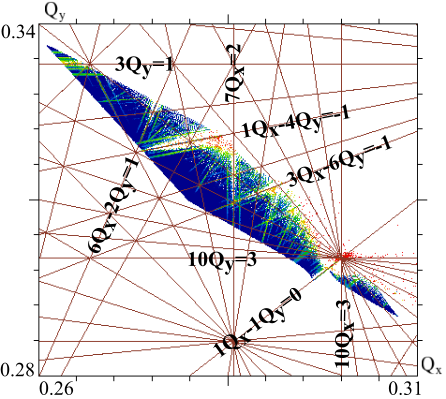} &
    \includegraphics[width=\fmawidth]{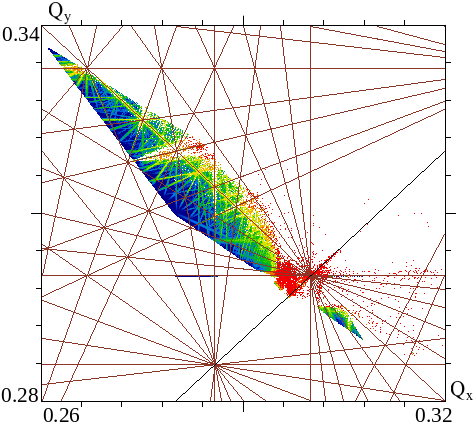}
    \\
    7th, H & 7th, V \\
    \includegraphics[width=\fmawidth]{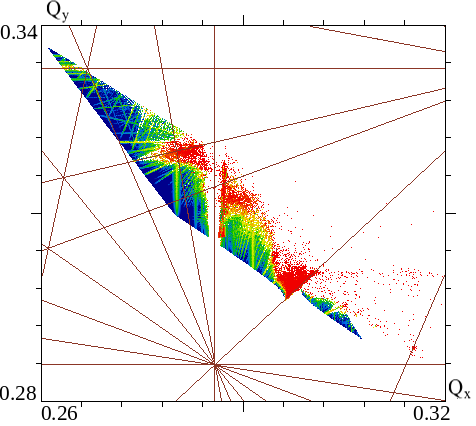} &
    \includegraphics[width=\fmawidth]{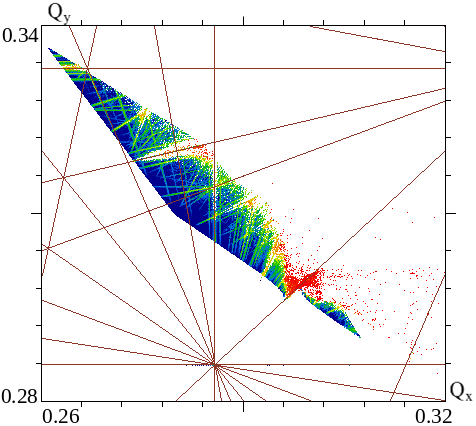} \\
  \end{tabular}
  \caption{Frequency-map analysis in betatron tune space based on the
    2016 injection optics with no machine errors and tunes (64.28,
    59.31): without excitation (top left); \tenthtp\ in both
    horizontal and vertical planes (top right); \seventhtp\ in the
    horizontal (bottom left) and vertical (bottom right) planes. The
    excitation amplitude is 120~nrad in the corresponding plane. The
    colors (blue to red) represent the tune jitter of tracked
    particles starting at each given location in tune
    space~\cite{fmalaskar}. The absence of a strong excitation of any
    resonance in case of vertical \seventhtp\ and the strong
    excitation in case of horizontal pulsing confirms the excitation
    of the $7 Q_x$ resonance. For \tenthtp, there is in contrast no
    significant difference between H, V, or H+V
    (Fig.~\ref{fig:fma:10}).}
  \label{fig:patternfma}
\end{figure*}

\begin{figure}
  \begin{tabular}{cc}
    No excitation & 10th, H+V \\
    \includegraphics[width=\smallfmawidth]{2016injnocolc15o+19_6noerru_dp0_ord10_annotate.png} &
    \includegraphics[width=\smallfmawidth]{2016injnocolc15o+19_6noerrut10skhv_dp0_ord10.png} \\
    10th, H & 10th, V \\
    \includegraphics[width=\smallfmawidth]{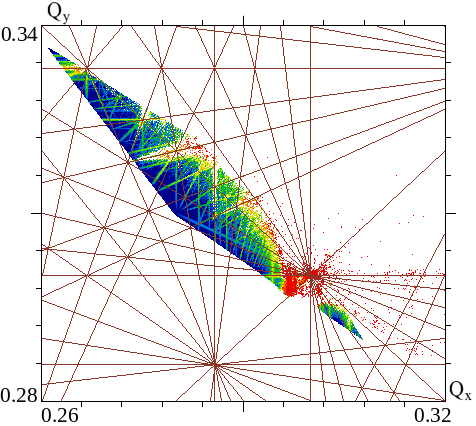} &
    \includegraphics[width=\smallfmawidth]{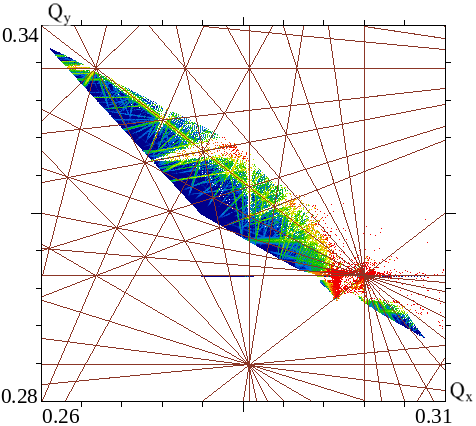} \\
  \end{tabular}
  \caption{FMA for \tenthtp\ based on the 2016 injection
    optics with no lattice errors and tunes (64.28, 59.31). The
    excitation is 120~nrad in the corresponding plane. There is no
    significant difference between pulsing in H only, V only, or in H+V.}
  \label{fig:fma:10}
\end{figure}

\begin{figure*}
  \begin{tabular}{cc}
    No excitation & Random, H+V \\
    \includegraphics[width=\bsrtwidth]{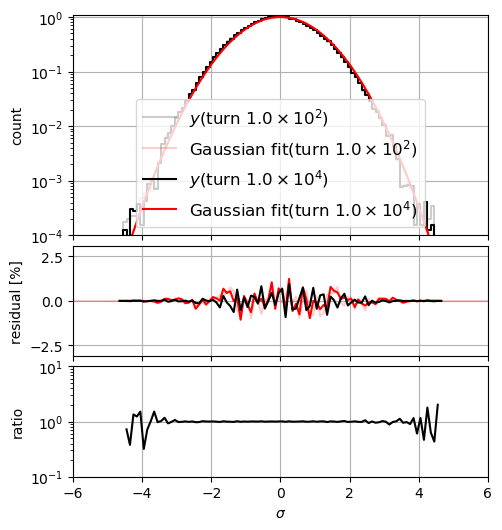} &
    \includegraphics[width=\bsrtwidth]{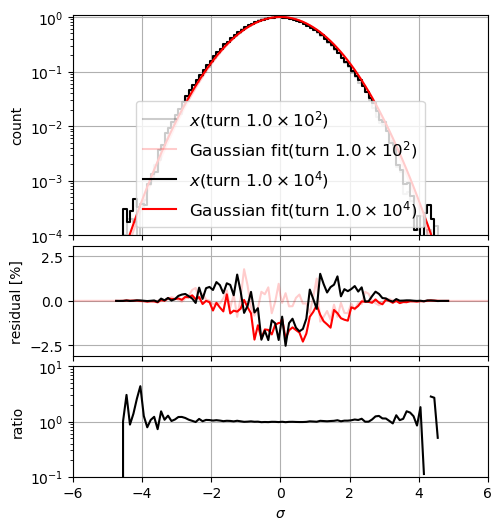} \\
    7th, H+V & 10th, H+V \\
    \includegraphics[width=\bsrtwidth]{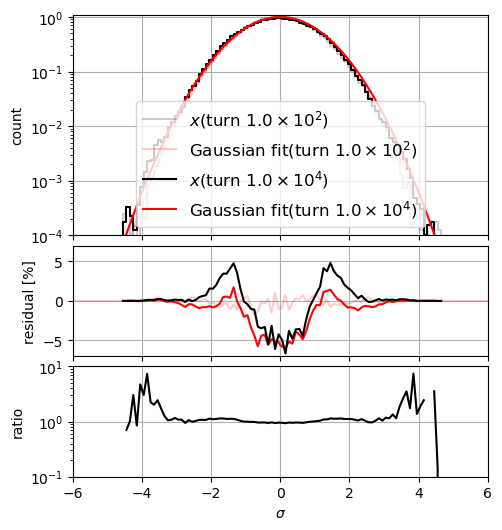}
                  &
    \includegraphics[width=\bsrtwidth]{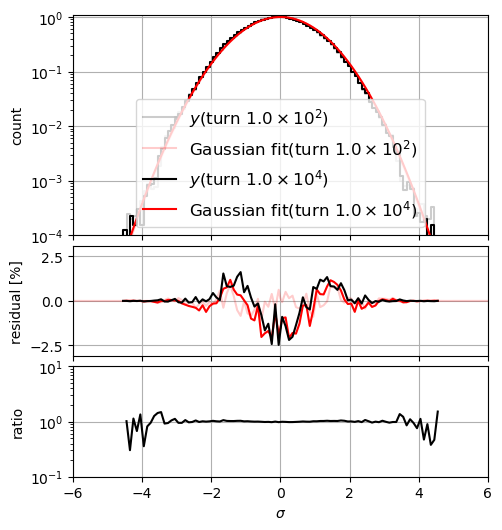} \\
  \end{tabular}
  \caption{Calculated beam distributions as a function of vertical
    position from distribution-tracking simulations based on the 2016
    injection optics with $(Q_x, Q_y) = (64.28, 59.31)$ and standard
    lattice errors: no excitation (top left), random excitation (top
    right), \seventhtp\ (bottom left), and
    \tenthtp\ (bottom right). The excitations
    are applied in both planes with an amplitude of 96~nrad. For each
    of the 4~cases, 3~plots are shown. The top plot shows the
    normalized transverse distributions: `initial' (after
    $10^2$~turns, in gray), `final' (after $10^4$~turns, in black),
    and their Gaussian fits (light and dark red, respectively). The
    middle plots show the relative residuals (i.e., differences, in
    percent of the peak value) between final and initial distributions
    (in black) and between each distribution and its Gaussian fit (in
    light and dark red). The ratios between final and initial
    distributions are drawn in black in the bottom plots. The
    residuals emphasize changes near the core of the distributions,
    whereas ratios (when statistically significant), reveal variations
    in the tails.}
  \label{fig:patternhist}
\end{figure*}

The effect of each pulsing pattern is characterized in terms of the
resulting losses and emittance growth, calculated in
distribution-tracking simulations.

As an example, the simulation results for the 2016 experiment are
shown in Fig.~\ref{fig:patternsim}.  A clear dependence of both losses
and emittances on the pattern can be seen.  The largest losses are
predicted for 3rd-, 7th- and \tenthtp, and for a uniform random
excitation. Because the bunch length decreases with the number of lost
particles, losses are attributed to off-momentum particles hitting the
transverse aperture. Significant emittance growth is visible only for
7th- and \tenthtp\ and for the random excitation. Compared to any
resonant excitation, the random excitation shows by far the strongest
effect.

The effects of 7th- and \tenthtp\ are also observed in the absence of
machine lattice errors. In this case, the only sources of nonlinearity
are sextupoles and octupoles~\cite{md_sim_hel_res_ex_fitterer},
suggesting that these nonlinearities are responsible for the
pronounced beam sensitivity. The driven resonances are revealed by the
frequency-map analysis, shown in Fig.~\ref{fig:patternfma}. The
$7 Q_x$ resonance is excited by the \seventhtp, and the $10 Q_x$ and
$10 Q_y$ resonances by the \tenthtp. As octupoles can only drive even
resonances, the sources of the $7 Q_x$ resonances are the sextupoles,
while the octupoles generate the tune footprint. The other pulsing
patterns do not exhibit any increase in losses or emittance growth
without magnetic errors. Their effect can thus be attributed to an
interaction of the excitation with the machine errors, implying also a
sensitivity to the chosen random seed in simulations and to the
specific error distribution in experiments.

In Fig.~\ref{fig:patternsim}, one can see that, in the cases of 7th-
and \tenthtp, the emittances start from an increased initial value and
then stay almost constant for the duration of the simulation. This
behavior is typical of the resonant excitation and it is not an
artifact of the simulation. It was verified that it is due to the
adjustment of the input beam distribution to a new equilibrium during
the first $10^4$~turns. In all long-term simulations ($10^6$~turns,
corresponding to 89~s) presented in this paper, the beam distribution
is saved every $10^4$~turns. Therefore, this initial fast adjustment
manifests itself in an increased initial emittance value.

Besides having a larger emittance, the new distribution also differs
from a Gaussian distribution.  This is illustrated in
Fig.~\ref{fig:patternhist}, which shows, for different excitation
patterns, two snapshots of the normalized transverse beam
distributions, the `initial' one at $10^2$~turns and the `final' one
at $10^4$~turns, and how they differ from each other and from a
Gaussian distribution. For instance, for \seventhtp\ (bottom left plot
in Fig.~\ref{fig:patternhist}), simulations indicate a clear shift in
beam population from the core ($< 1\sigma$) towards the regions at
$\pm 1.5\sigma$ and towards the tails at $\pm 4\sigma$.

Although on a different time scale (minutes instead of seconds), a
similar re-adjustment of the beam distribution was directly observed
for the first time, to our knowledge, during the 2016 and 2017 LHC
experiments presented in this paper, as discussed below.

If only the resonant excitation is present, without noise, simulations
show that the modified distribution is stable, with constant
emittance. By adding a random noise component, representative of the
natural noise present in the LHC, one calculates, after the initial
adjustment phase, emittance-growth rates that increase with the
amplitude of the excitation. As discussed below, an increase of the
emittance-growth rates with the applied excitation amplitude was also
observed experimentally.

In the case of random excitation, there is constant emittance growth,
without any initial adjustment phase. The interaction with the natural
noise sources in the machine results in increased losses and emittance
growth, as it is basically equivalent to the application of a random
excitation with an increased amplitude. In addition to the experiments
presented in this paper, the effect of a random excitation was also
studied in a separate experiment at the LHC for the case of colliding
beams~\cite{md1433_noise_top_energy, md_noise_bbLHC}.

\subsection{Pulsing every 10th turn}
\label{sec:simex10}

\begin{figure*}
  \begin{tabular}{ccc}
    \includegraphics[width=\thirdwidth]{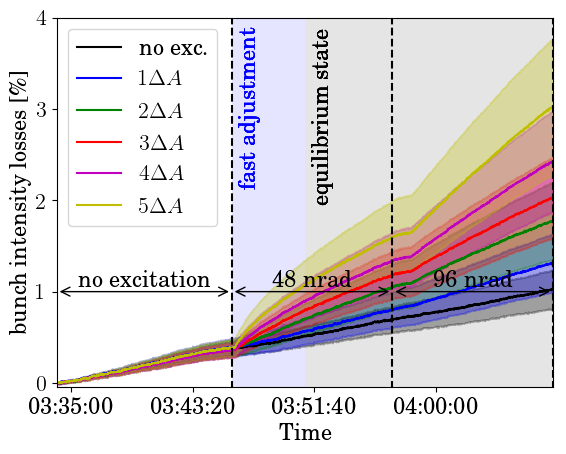}&
    \includegraphics[width=\thirdwidth]{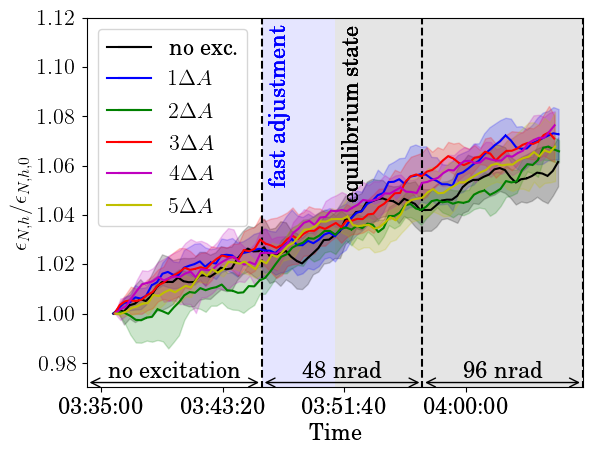}&
    \includegraphics[width=\thirdwidth]{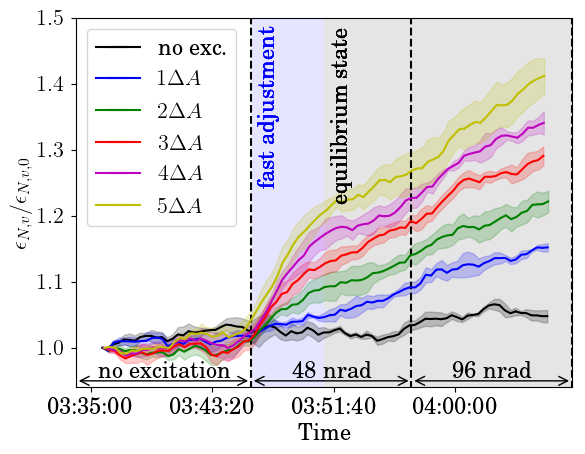}\\
  \end{tabular}
  \caption{Summary of the 2016 experiments on \tenthtp\ in the
    vertical plane: losses (left), horizontal emittances (middle) and
    vertical emittances (right), relative to their initial values. The
    transverse damping system was not active in this case. The 3
    excitation periods are labeled in black according to the value of
    the maximum excitation amplitude
    $A_{\mathrm{max}} = 5 \, \Delta A$: no excitation, 48~nrad or
    96~nrad. The 4~bunches experiencing the same excitation amplitude
    $n \, \Delta A$ ($n = 0, \ldots, 5$) are grouped by color. The
    data are averaged over the 4~bunches, with the envelope
    representing the standard deviation. The area with a blue
    background highlights qualitatively the fast adjustment period of
    the beam distribution, transitioning into a new equilibrium state
    (indicated by the gray background).}
  \label{fig:10thexp}
\end{figure*}

\begin{figure*}
  \begin{tabular}{cc}
    Reference bunch, no excitation & Excited bunch, maximum amplitude \\
    \includegraphics[width=\bsrtwidth]{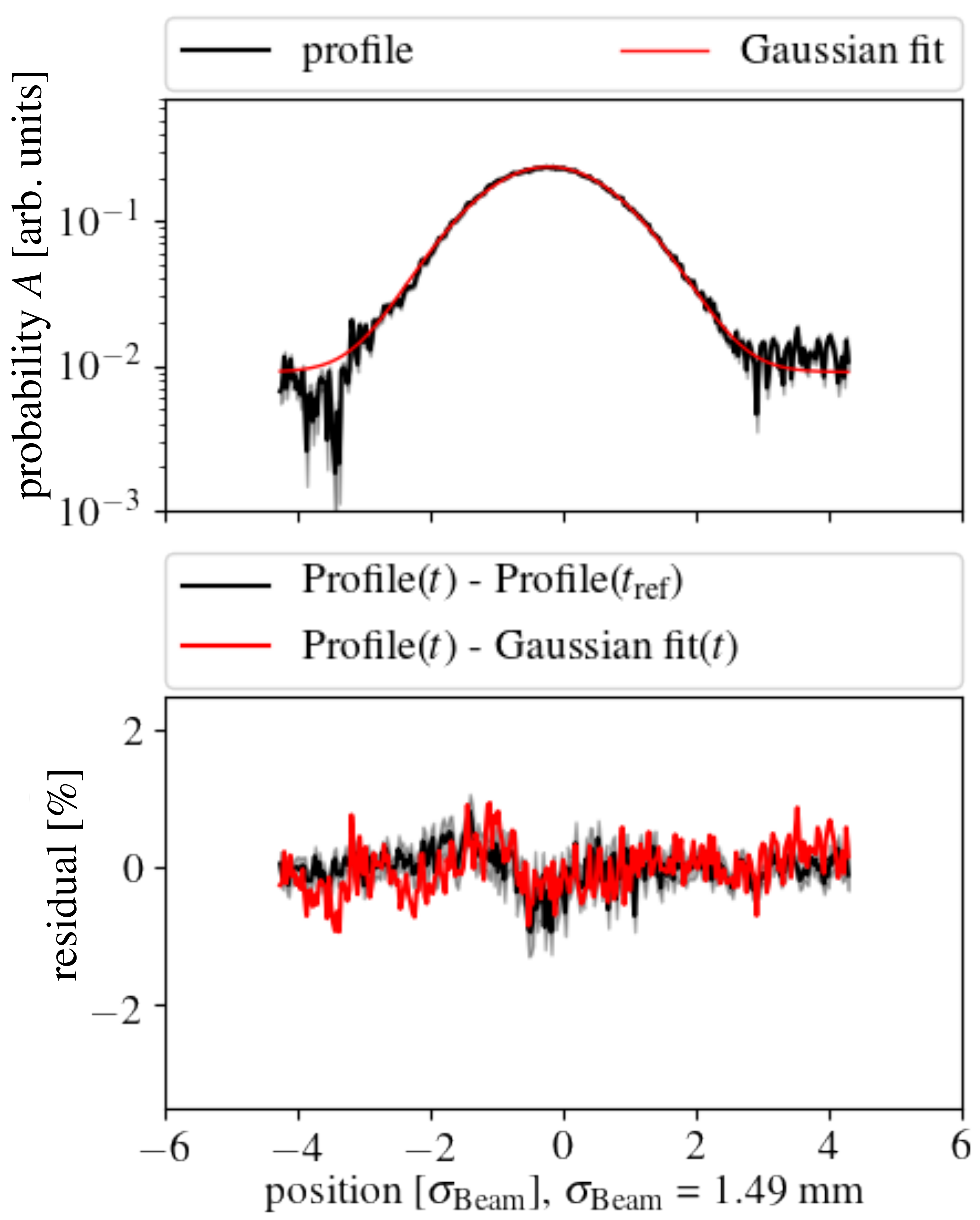} &
    \includegraphics[width=\bsrtwidth]{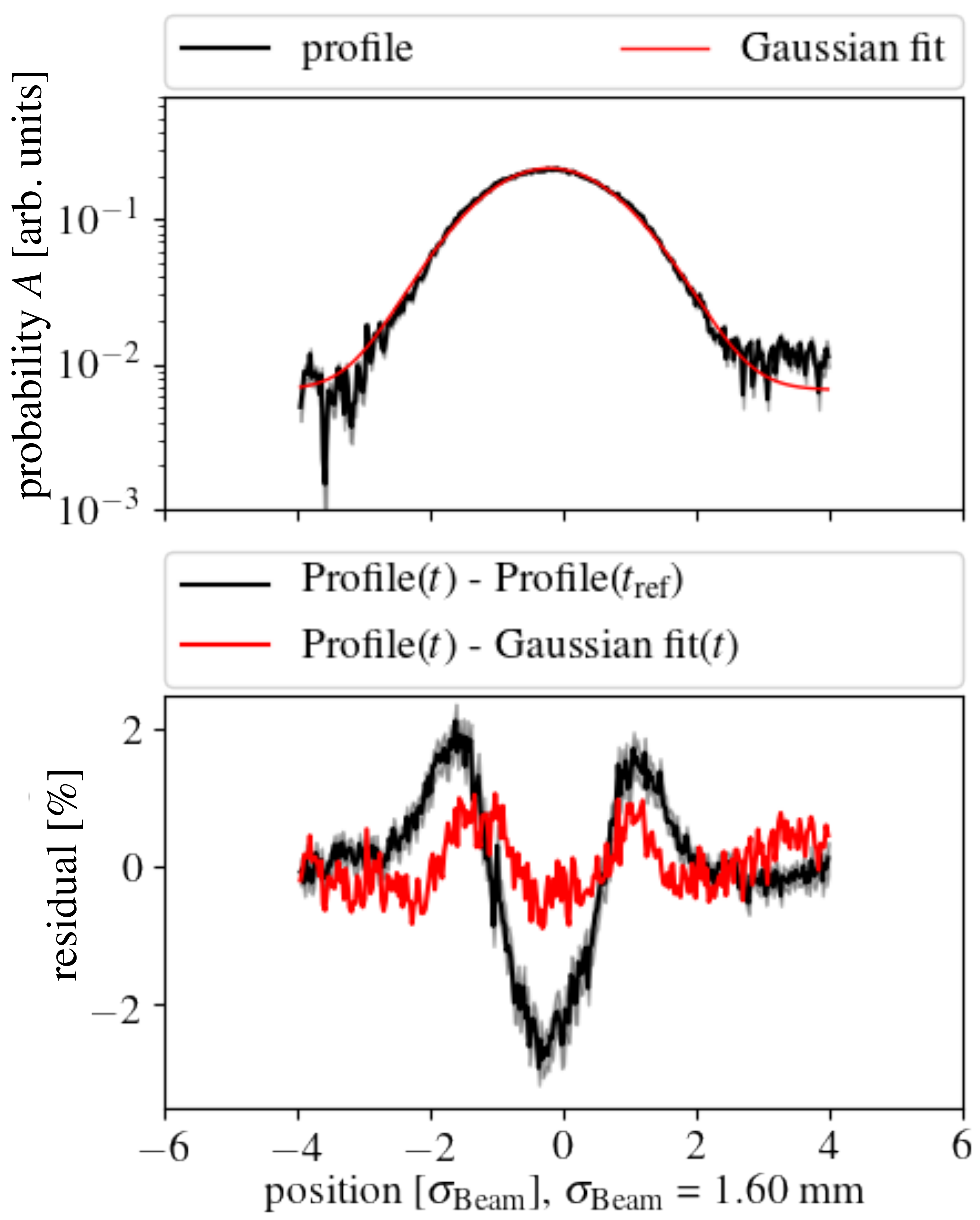}\\
    \end{tabular}
    \caption{Vertical beam profiles measured with the Beam Synchrotron
      Radiation Telescope (BSRT) during the 2016 experiments on
      \tenthtp\ in the vertical plane. The
      transverse damping system was not active on these bunches. The
      beam distributions at the end of the excitation period are shown
      in black in the top plots, with a Gaussian fit in red. The
      bottom plots show the residuals: final profile minus initial
      profile (black); final profile minus its Gaussian fit
      (red). Residuals are expressed as a fraction of the peak profile
      amplitude. The black lines in the plots of the residuals are a
      measure of the overall change of the distribution shape. The red
      lines indicate how different the final distributions are from a
      Gaussian shape. Details of the analysis are given in
      Ref.~\cite{bsrtprofinj}. The distribution of the reference bunch
      (left) is almost unchanged, whereas the bunch experiencing the
      maximum excitation (right) shows a clear shift of particles from
      the axis towards approximately $\pm 2\sigma$.}
    \label{fig:10thexpprof}
\end{figure*}

\begin{figure*}
  \begin{tabular}{cc}
    \includegraphics[width=\halfwidth]{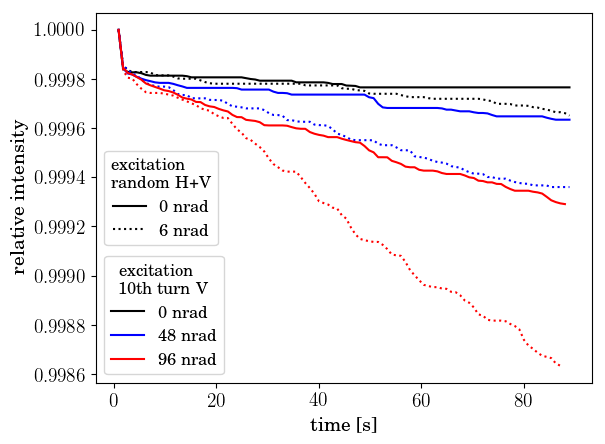} &
    \includegraphics[width=\halfwidth]{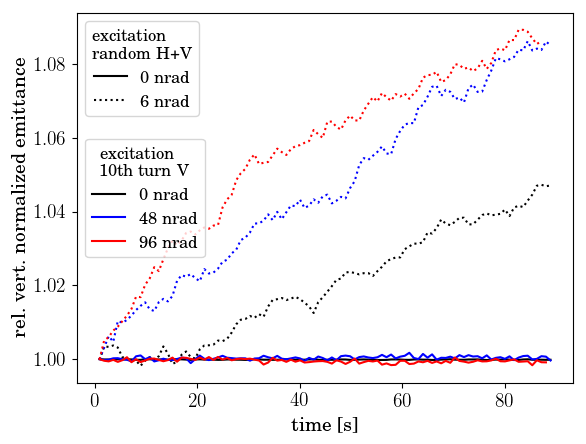} \\
    \includegraphics[width=\halfwidth]{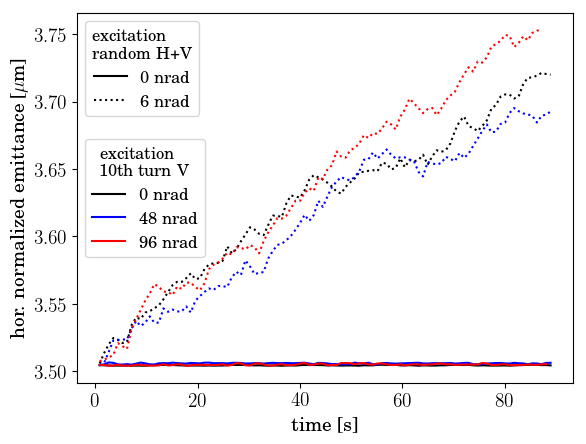} &
    \includegraphics[width=\halfwidth]{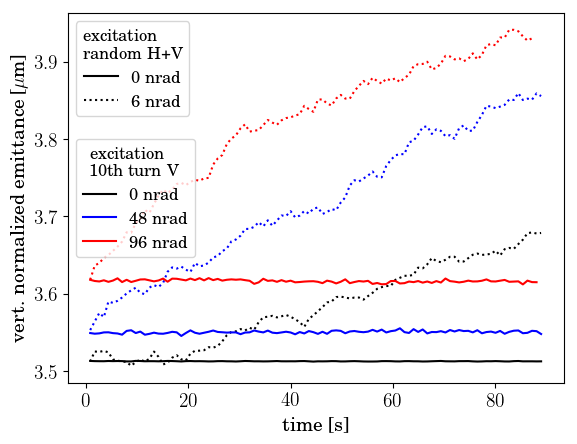} \\
  \end{tabular}
  \caption{Simulations (with the distribution-tracking method) based
    on the 2016 injection optics with standard lattice errors and
    tunes $(Q_x, Q_y) = (64.28, 59.31)$: relative beam intensity (top
    left); normalized horizontal (bottom left) and vertical (bottom
    right) emittances; vertical emittance relative to its initial
    value (top right). The solid lines indicate the effect of
    \tenthtp\ in the vertical plane. The dotted
    lines include both the resonant excitation and a random dipole
    noise component of 6~nrad (in both horizontal and vertical
    planes).}
  \label{fig:10thsim}
\end{figure*}

\begin{figure*}
  \centering
  \includegraphics[width=\twothirdswidth]{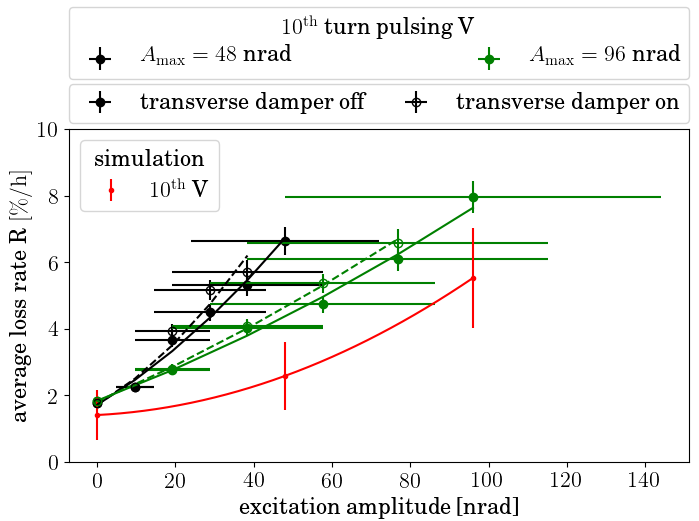}
  \caption{Comparison between experiments and simulations of loss
    rates vs.\ resonant excitation amplitude for
    \tenthtp\ in the vertical plane. The
    relative average loss rate~$R$ is defined in
    Eq.~\ref{eqn:lossrate}. The experimental results for a maximum
    excitation amplitude of 48~nrad are shown in black and those for
    96~nrad are plotted in green (see also Fig.~\ref{fig:10thexp},
    left, for instance). The results with damper off are represented
    by solid dots, those with damper on are shown with open
    circles. The simulation results, including random dipole noise,
    are shown in red (see also Fig.~\ref{fig:10thsim}). Statistical
    and systematic uncertainties are discussed in the text. The curves
    represent empirical quadratic fits to the data with damper off
    (solid) and with damper on (dashed).}
  \label{fig:10thexploss}
\end{figure*}

The \tenthtp\ pattern was tested in 2016 with an
excitation in the vertical plane only. As this was the first time a
resonant excitation was tested in the LHC, only 48~bunches were
injected, in order to guarantee safe operation of the machine. To test
different excitation amplitudes during one fill and have enough
statistics, the excitation was only applied in one plane. The
experiment started with a period of 12~minutes without excitation, to
let the beam distribution fully adjust to its equilibrium state after
injection. The excitation was then applied for 11~minutes, with the
excitation scheme shown in Fig.~\ref{fig:fill} and a maximum
excitation amplitude of
$A_{\mathrm{max}} = 5 \, \Delta A = \q{48}{nrad}$. The excitation was
then further increased to the maximum value of
$A_{\mathrm{max}} = 5 \, \Delta A = \q{96}{nrad}$ and kept for another
period of 11~minutes.

The main observations are collected in Figures~\ref{fig:10thexp}
and~\ref{fig:10thexpprof}. In Fig.~\ref{fig:10thexp}, one can see the
evolution in time of losses and emittances for the control bunches and
for the affected bunches, as a function of excitation amplitude. In
brief, during both phases of the experiment, the excitation induced
the following changes:
\begin{itemize}
  \item loss rates increased with excitation amplitude
    (Fig.~\ref{fig:10thexp}, left);
  \item emittance growth increased with amplitude in the vertical
    plane, but not in the horizontal plane (Fig.~\ref{fig:10thexp},
    center and right);
  \item a change of the beam distribution for the excited bunches was
    directly observed (Fig.~\ref{fig:10thexpprof}).
\end{itemize}

As loss rates, emittance-growth rates and beam distributions of the 4
reference bunches were unchanged, the above observations can be
directly attributed to the pulsing pattern. The quantitative
dependence of the loss rates and emittance-growth rates on excitation
amplitude measured in these experiments provides an estimate of the
effects of external resonant excitations, such as the residual fields
of a pulsed hollow electron lens.

The vertical emittance features a behavior similar to that predicted
in simulations: a fast adjustment phase of the beam distribution,
which manifests itself as a rapid increase of the emittance, followed
by a new equilibrium, inferred from a slower and continuous emittance
growth. In Fig.~\ref{fig:10thexp}, these two phases are indicated in
blue and black.

Distribution changes could also be directly observed, thanks to the
performance of the Beam Synchrotron Radiation Telescope
(BSRT)~\cite{Trad:PhD:2015, bsrtprofinj}. As an example,
Fig.~\ref{fig:10thexpprof} shows the vertical profile of a reference
bunch and of a bunch experiencing the maximum excitation
$A_{\mathrm{max}} = 5 \, \Delta A = \q{48}{nrad}$ and later
96~nrad. While the reference bunch stays unchanged, the distribution
of the excited bunch clearly changes and re-adjusts to a non-Gaussian
shape, with particles shifting from the core towards the regions at
approximately $\pm 2\sigma$. This behavior is similar to what was
calculated in simulations, as discussed in Section~\ref{sec:pattern}
and in Fig.~\ref{fig:patternhist}.

Numerical simulations of intensities and emittances for this
experiment are shown in Fig.~\ref{fig:10thsim}. They are the results
of distribution tracking for \tenthtp\ without
additional noise (solid lines) and with an additional noise component
(dotted lines), to emulate the natural noise present in the LHC
(Eq.~\ref{eq:noise-kick}). The presence of noise significantly changes
the calculated effect of the resonant excitation on both losses and
emittances.

A direct comparison of measured and predicted loss rates is presented
in Fig.~\ref{fig:10thexploss}. As experiments and simulations were
conducted on different time intervals (11~minutes vs.\ 90~s), we
compare the relative average loss rate~$R$:
\begin{equation}
  \label{eqn:lossrate}
  R = \frac{I_{\mathrm{start}} - I_{\mathrm{end}}}{I_{\mathrm{start}}}
  \cdot \frac{1}{\Delta t},
\end{equation}
where $I$ is the beam intensity and
$\Delta t = t_\mathrm{end} - t_\mathrm{start}$ is the time interval
during which the excitation with given amplitude is
applied. Uncertainties on the excitation amplitude arise from the
calibration of the transverse damping system
(Section~\ref{sec:adt}). For both experiments and simulations, the
error bars on the loss rate are statistical. The magnitude of the
random noise component introduces a systematic uncertainty in the
calculations, which was not evaluated for the scope of this paper. The
systematic uncertainty on the measurements can be inferred from the
difference between the 2~periods of the experiment, which should in
principle be the same. Part of this systematic effect could be due to
the fact that the second set of excitations with double amplitude
started with a modified beam distribution. The magnitude of this
effect could be verified in future experiments in which the
excitations are reversed or where a new beam fill is used for each set
of measurements. Overall, we conclude that the effects of this
resonant excitation on loss rates could be measured in the LHC within
a factor~2 and could be predicted with similar accuracy.

\subsection{Pulsing every 7th turn}
\label{sec:simex7}

\begin{figure*}
  \begin{tabular}{cc}
    no excitation & 7th, H+V \\
    \includegraphics[width=\fmawidth]{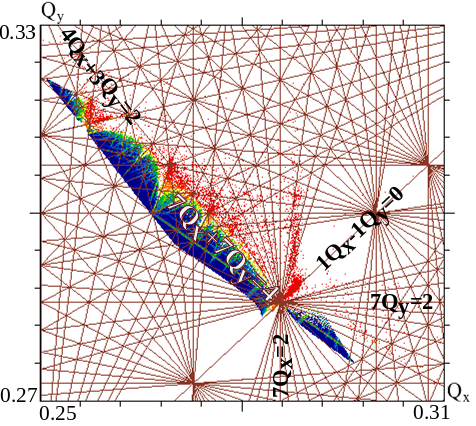} &
    \includegraphics[width=\fmawidth]{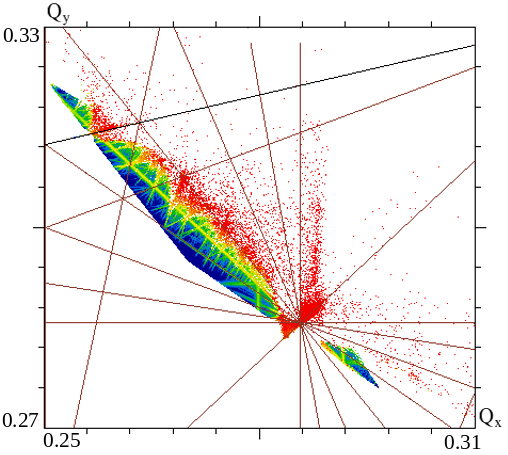} \\
    7th, H & 7th, V \\
    \includegraphics[width=\fmawidth]{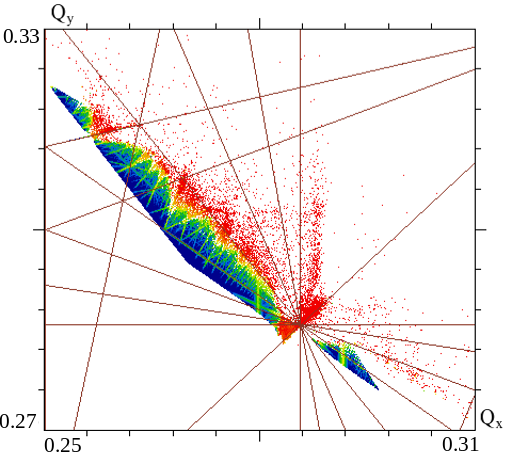} &
    \includegraphics[width=\fmawidth]{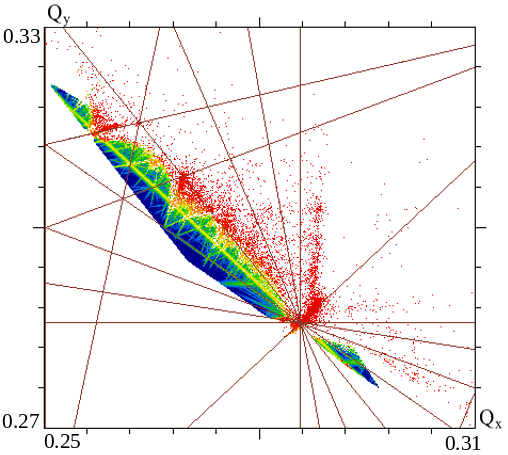} \\
  \end{tabular}
  \caption{FMA in betatron tune space based on the 2017 injection
    optics with no machine errors and tunes (62.27, 60.295): without
    excitation (top left) and for \seventhtp\ in H+V (top right), H
    only (bottom left) and V only (bottom right). The excitation
    amplitude is 96~nrad. The $7 Q_x$, $7 Q_y$, and 14th-order
    $7 Q_x + 7 Q_y$ resonances are excited.}
  \label{fig:7th2017fma}
\end{figure*}

\begin{figure*}
  \begin{tabular}{cc}
    \includegraphics[width=\halfwidth]{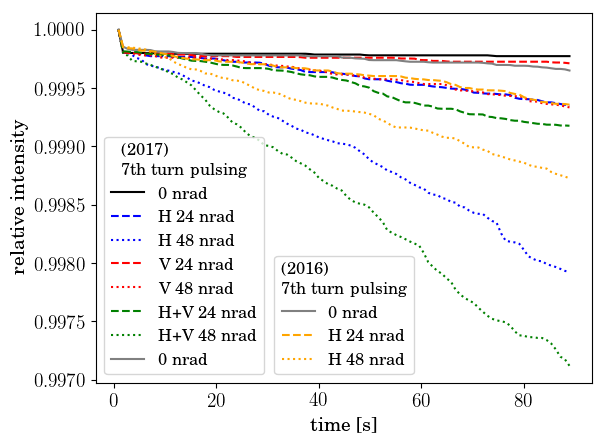} &
    \includegraphics[width=\halfwidth]{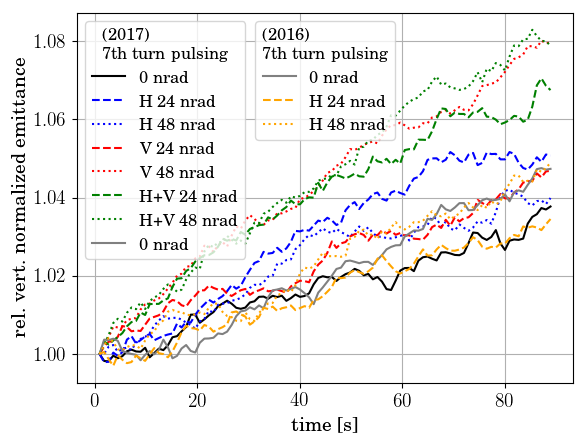} \\
    \includegraphics[width=\halfwidth]{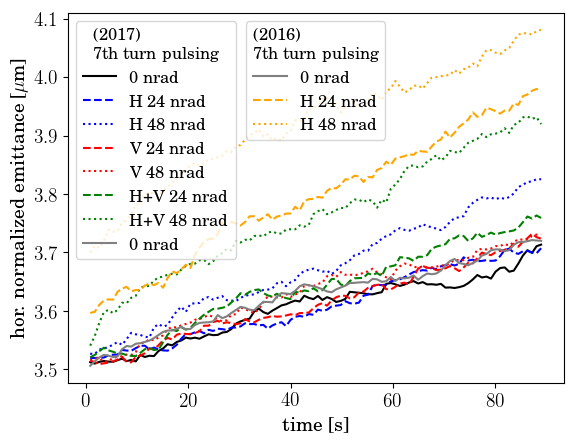} &
    \includegraphics[width=\halfwidth]{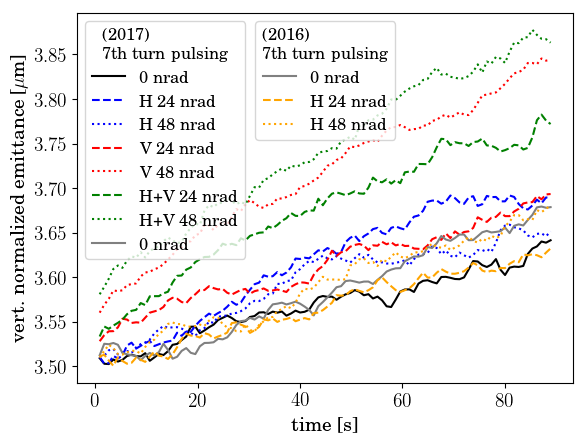} \\
  \end{tabular}
  \caption{Calculated bunch intensities and emittances from
    distribution tracking based on the 2016 injection optics with
    standard lattice errors and $(Q_x, Q_y) = (64.28, 59.31)$ and 2017
    injection optics with standard lattice errors and
    $(Q_x, Q_y) = (62.27, 60.295)$: relative intensity (top left);
    horizontal emittance (bottom left); vertical emittance (bottom
    right); the relative vertical emittance is also shown (top
    right). The solid black line includes only a random dipole noise
    component in H+V of 6~nrad. The dotted and dashed lines correspond
    to \seventhtp\ with two different excitation amplitudes (24~nrad
    and 48~nrad), plus a random dipole noise component in H+V of
    6~nrad.}
  \label{fig:7thsim}
\end{figure*}

\begin{figure*}
  \begin{tabular}{ccc}
    \multicolumn{3}{c}{2016 experiment, 7th, H}	\\
    \includegraphics[width=\thirdwidth]{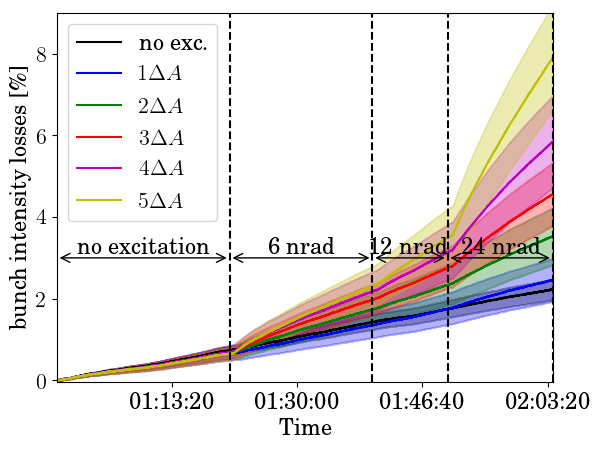} &
    \includegraphics[width=\thirdwidth]{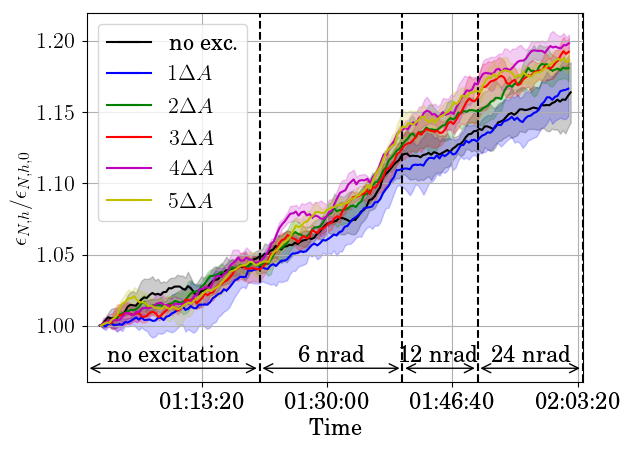} &
    \includegraphics[width=\thirdwidth]{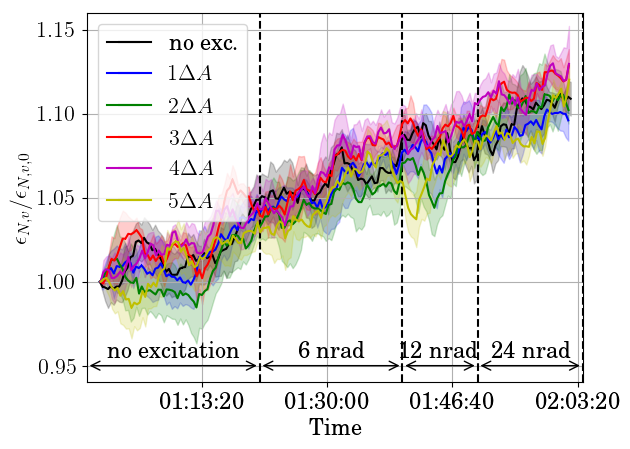} \\
    \multicolumn{3}{c}{2017 experiment, 7th, H+V} \\
    \includegraphics[width=\thirdwidth]{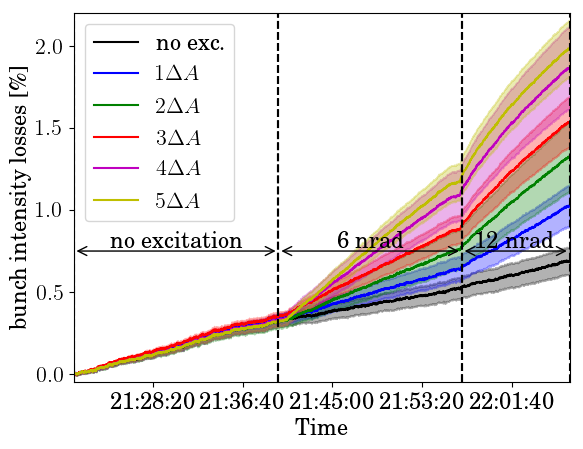} &
    \includegraphics[width=\thirdwidth]{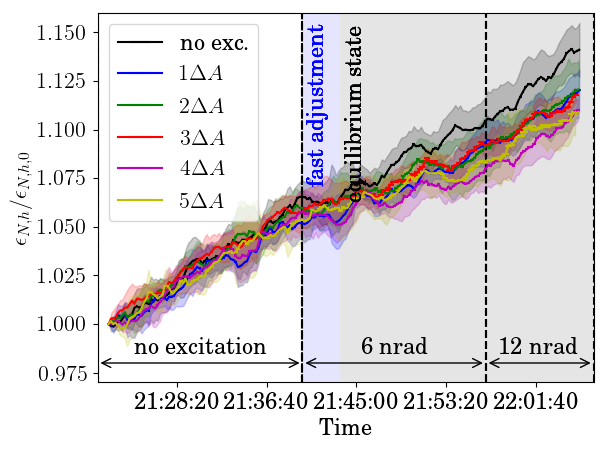} &
    \includegraphics[width=\thirdwidth]{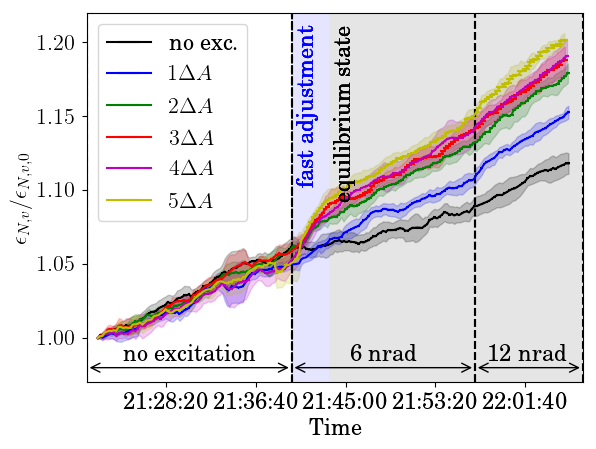} \\
  \end{tabular}
  \caption{Measured losses and emittances during the 2016 and 2017
    experiments: relative losses (left), relative horizontal
    emittances (middle), and relative vertical emittances
    (right). Measurements are averaged over the bunches experiencing
    the same excitation amplitude. The transverse damping system was
    not active in this set of measurements.}
  \label{fig:7thexp}
\end{figure*}

\begin{figure*}
  \centering
  \includegraphics[width=\twothirdswidth]{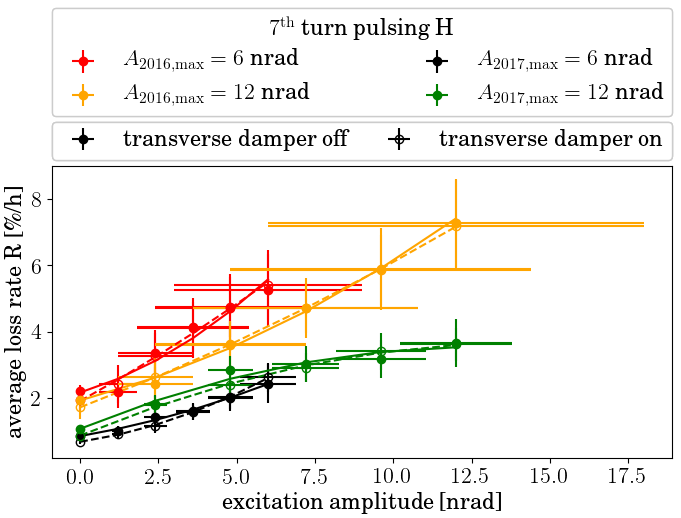}
  \caption{Measured loss rates vs.\ excitation amplitude for
    \seventhtp\ in the horizontal plane during the 2016 experiment
    (red and yellow) and the 2017 experiment (black and green). The
    uncertainties on the excitation amplitude are dominated by the
    calibration of the transverse feedback and damping system. The
    uncertainties on the loss rates are statistical. An estimate of
    the systematic uncertainties (due to changes in beam distribution,
    for instance) is given by the difference between the two data sets
    within each experiment. The lines indicate empirical second-order
    polynomial fits, with damper off (solid) and with damper on
    (dashed).}
  \label{fig:7thexploss}
\end{figure*}

\begin{figure*}
  \begin{tabular}{c}
    \includegraphics[height=\onethirdheight]{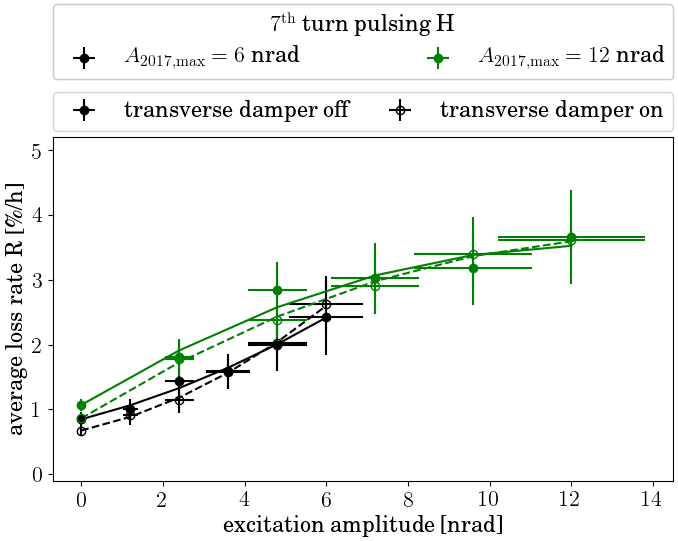} \\
    \includegraphics[height=\onethirdheight]{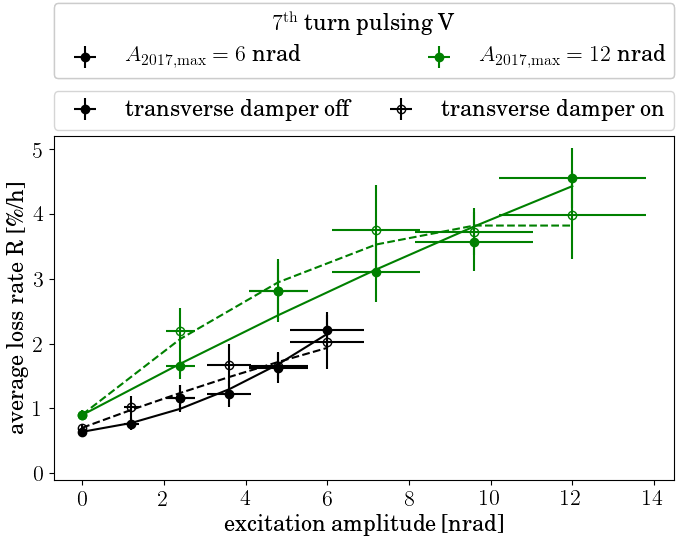} \\
    \includegraphics[height=\onethirdheight]{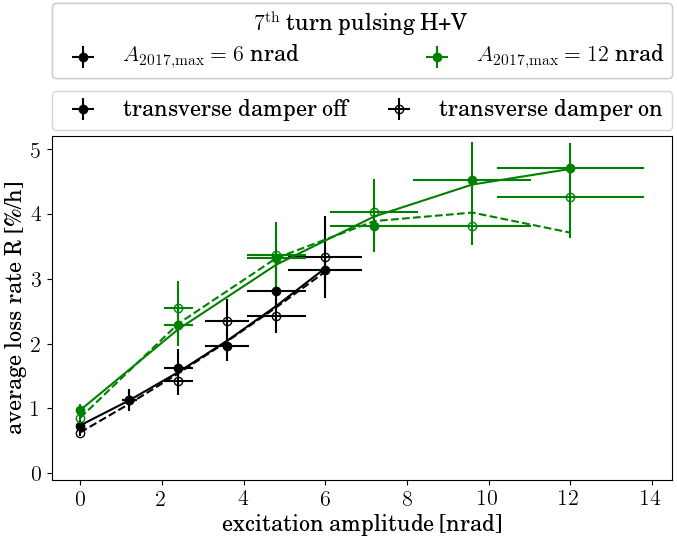} \\
  \end{tabular}
  \caption{Comparison of the measured loss rates as a function of
    excitation amplitude during the 2017 experiment for \seventhtp\ in
    H only (left), V only (center), and in H+V (right). The three
    excitations generate similar loss rates.}
  \label{fig:7thexploss2017}
\end{figure*}

\begin{figure*}
  \begin{tabular}{cc}
    Reference bunch, no excitation & Excited bunch, 7th, H+V \\
    \includegraphics[width=\bsrtwidth]{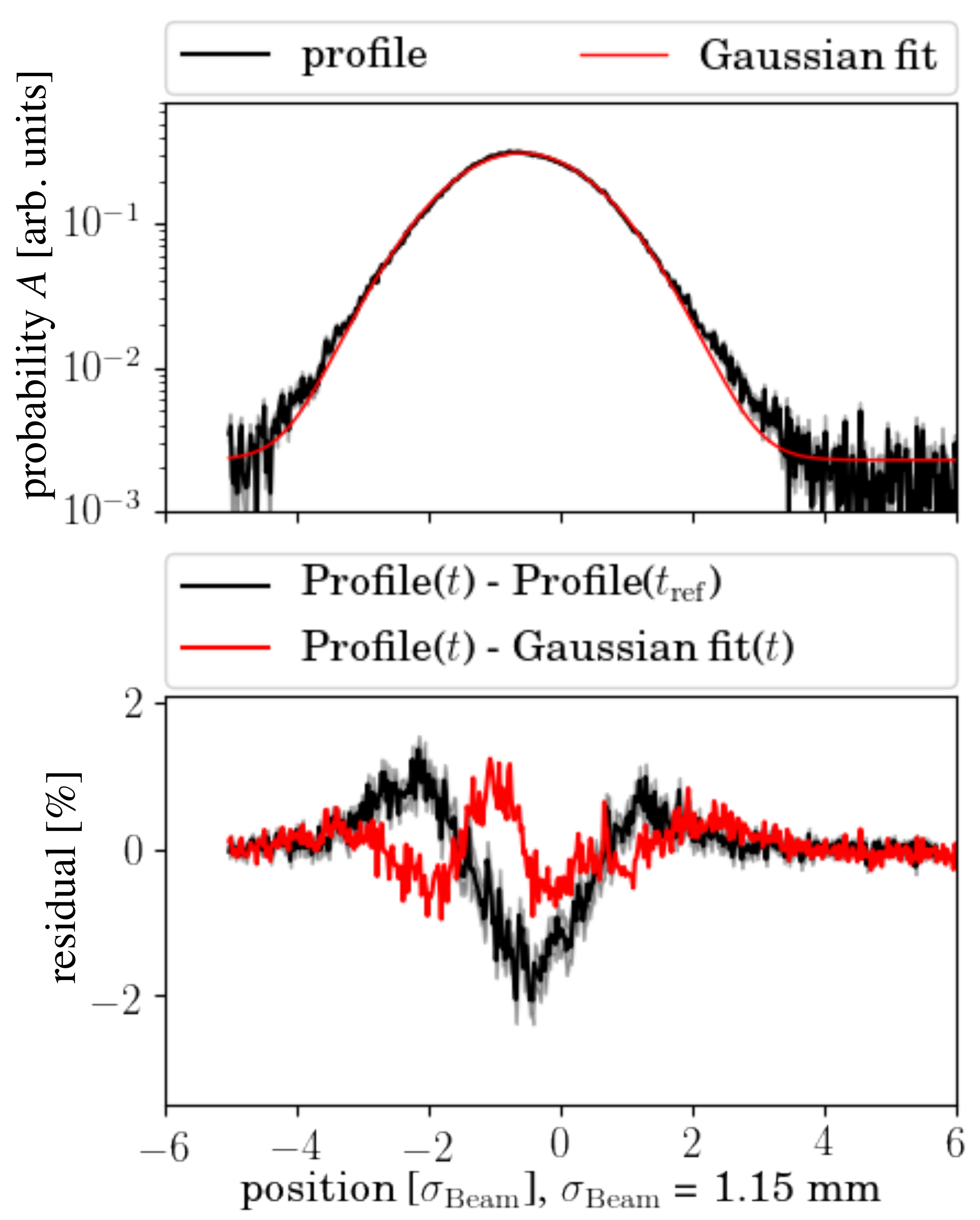} &
    \includegraphics[width=\bsrtwidth]{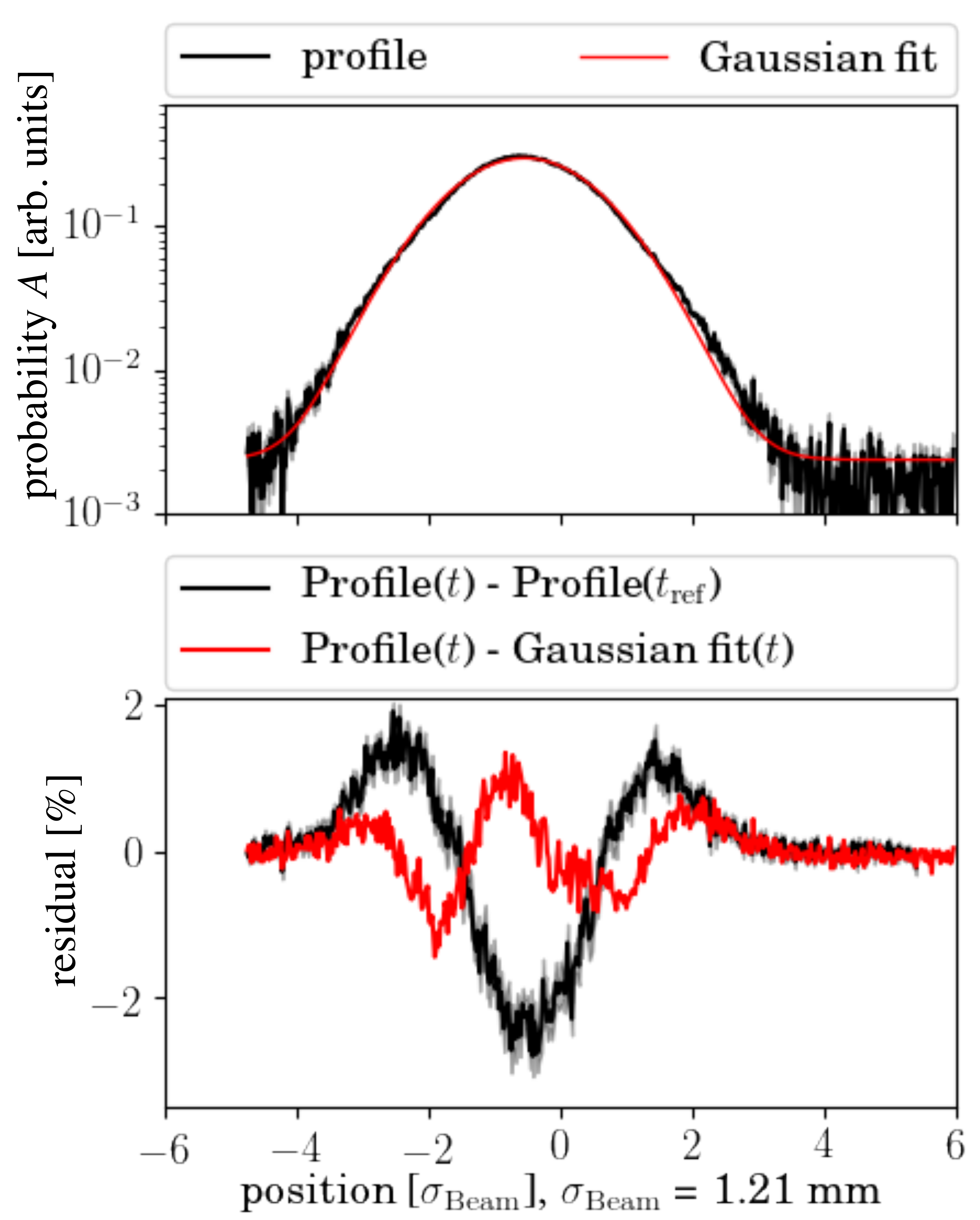} \\
  \end{tabular}
  \caption{Vertical beam profiles measured with the Beam Synchrotron
    Radiation Telescope (BSRT) during the 2017 experiments. The
    profiles are taken at the end of \seventhtp\ in H+V. For these
    bunches, the transverse damper was not active. The data are
    presented in the same way as in Fig.~\ref{fig:10thexpprof}. The
    profile changes in the bunch affected by the maximum excitation
    (right) are larger than those displayed by the reference bunch
    (left).}
  \label{fig:7thexpprof}
\end{figure*}

The \seventhtp\ pattern was tested during both the
2016 and 2017 experiments. In 2016, the resonant excitation was
employed for the first time in the LHC and the experiments were
therefore still in an exploratory stage. Based on the experience
gained in 2016, the experiments were then repeated more
systematically, with more bunches and on both excitation planes.

The experiments were divided into different periods: the first period
without excitation; the second period with a resonant excitation with
a maximum amplitude of
$A_{\mathrm{max}} = 5 \, \Delta A = \q{6}{nrad}$; and a third period
in which the excitation amplitude was further increased to
$A_{\mathrm{max}} = \q{12}{nrad}$. In~2016, experiments included a
fourth period at $A_{\mathrm{max}} = \q{24}{nrad}$.

The machine tunes were changed in standard operations from (64.28,
59.31) in~2016 to (62.27, 60.295) in~2017. This change was accompanied
by a small change in optics, considered negligible for these
measurements. A direct comparison of the two experiments is therefore
not possible. However, these differences show how the effects of
resonant excitations are affected by small changes in fractional tune.

An insight on how the change in tune entails a change in driving
resonances is provided by frequency-map analysis.
Figure~\ref{fig:patternfma} (bottom left and bottom right plots),
which was described previously, shows a strong increase in diffusion
for \seventhtp\ in the horizontal plane around the $7Q_x$ resonances,
and only small changes for pulsing in the vertical plane. In the case
of the 2017 optics and tune, both the $7Q_x$ and the $7Q_y$ resonances
cross the tune footprint (Fig.~\ref{fig:7th2017fma}, top left). Both
of these resonances are excited by \seventhtp, as one can see from the
increase in tune jitter around the corresponding resonance lines for
each excitation plane separately, and thus also for their combined
action (Fig.~\ref{fig:7th2017fma}, bottom left, bottom right, and top
right plots). Although much weaker, an increase in tune diffusion is
also observed around the $7Q_x + 7Q_y$ line when pulsing in the
vertical plane or in both planes (diagonal yellow-green line in
Fig.~\ref{fig:7th2017fma}, top right and bottom right plots). This
suggests an excitation of the 14th-order resonance. The reason why
this phenomenon is much less pronounced for horizontal excitations is
not understood.

The simulated losses and emittances for both 2016 and 2017 conditions
are shown in Fig.~\ref{fig:7thsim}. The experimental results are
summarized in Figures~\ref{fig:7thexp}, \ref{fig:7thexploss},
\ref{fig:7thexploss2017}, and~\ref{fig:7thexpprof}.

Calculated losses for~2016 and~2017 are similar at 24~nrad and
increase more rapidly with amplitude for 2017 conditions
(Fig.~\ref{fig:7thsim}, top left). Horizontal pulsing is a few times
stronger than vertical pulsing, and the combined H+V action appears
slightly more effective than the sum of H and V separately.

Experimental losses are plotted as a function of time in
Fig.~\ref{fig:7thexp} and as a function of excitation amplitude in
Figs.~\ref{fig:7thexploss}
and~\ref{fig:7thexploss2017}. Figure~\ref{fig:7thexploss} shows that
measured losses were actually larger in~2016 than they were
in~2017. They were also a few times larger than those predicted by
simulations. Another discrepancy was observed in the excitation plane:
horizontal, vertical, and combined pulsing had similar effects
(Fig.~\ref{fig:7thexploss2017}). The same considerations on systematic
effects discussed in Section~\ref{sec:simex10} apply in this case.

This pulsing pattern generated stronger losses than expected. In this
case, we conclude that predictions are difficult, due to their
sensitivity to lattice configuration, noise sources, and beam
distributions.

The simulated emittances from distribution tracking are shown in
Fig.~\ref{fig:7thsim} (bottom left and right). The main calculated
results are the following, for the two experimental conditions:
\begin{description}
\item[2016 conditions] In this case, excitations were only in the
  horizontal plane. There is a fast increase of the horizontal
  emittance, dependent on the excitation amplitude, after which a
  constant emittance growth is observed. Vertical emittance growth is
  almost negligible.
\item[2017 conditions] Horizontal emittances are mostly affected by
  the amplitude of H+V excitations and, to a much smaller extent, by
  H~excitations; no effect is observed from V excitations. Vertical
  emittances are mostly affected by the amplitude of H+V and V
  excitations; the effect of H excitations is much smaller, if any,
  with no clear dependence on amplitude.
\end{description}

A set of measured emittances is plotted in Fig.~\ref{fig:7thexp}
(middle and right plots). In~2016, with only horizontal excitations,
amplitude-dependent horizontal emittance growth was observed, with no
effect on the vertical emittance. In~2017, V and H+V excitations
generated vertical emittance growth, while horizontal excitations had
no effect. These observations are consistent with the predicted
behavior.

During these experiments, emittances changed in two phases, as they
did during the \tenthtp\ studies
(Section~\ref{sec:simex10}) and in simulations: a fast adjustment
phase (shaded in blue in Fig.~\ref{fig:7thexp}) followed by a new
equilibrium (in gray). Changes in the transverse beam distribution
could be observed directly through changes in the synchrotron-light
(BSRT) profiles (Fig.~\ref{fig:7thexpprof}).

\subsection{Pulsing every 8th turn}
\label{sec:simex8}

\begin{figure*}
  \begin{tabular}{cc}
  \includegraphics[width=\halfwidth]{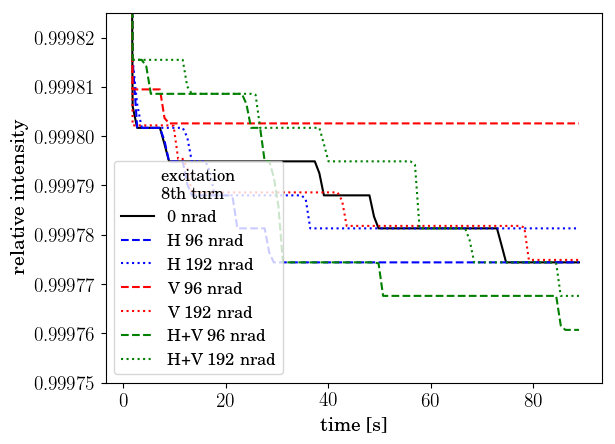} &
  \includegraphics[width=\halfwidth]{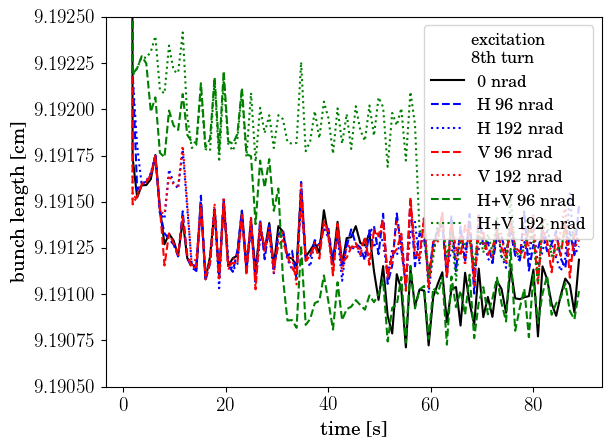} \\
  \includegraphics[width=\halfwidth]{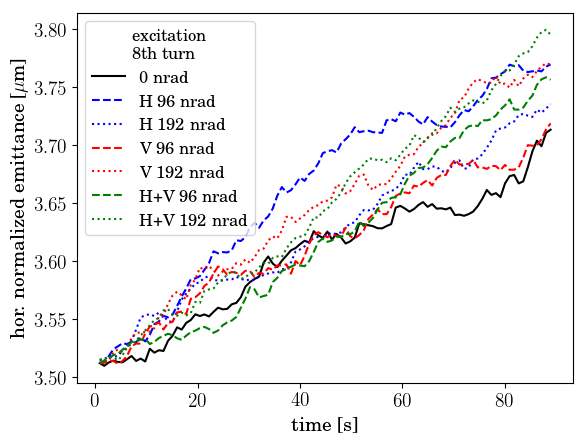} &
  \includegraphics[width=\halfwidth]{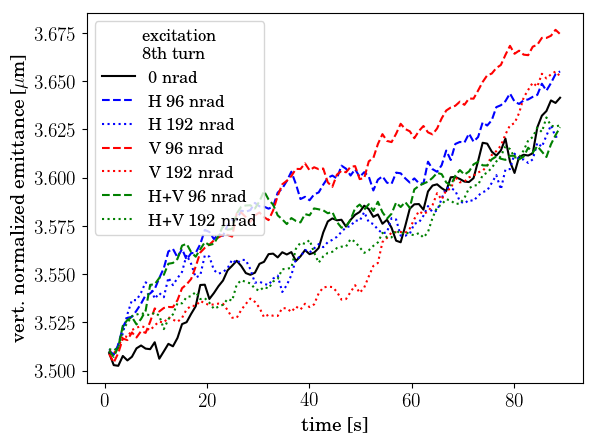} \\
  \end{tabular}
  \caption{Calculated relative intensities (top left), bunch lengths
    (top right), horizontal emittances (bottom left), and vertical
    emittances (bottom right) from distribution-tracking simulations
    based on the 2017 injection optics with standard lattice errors
    and tunes $(Q_x, Q_y) = (62.27, 60.295)$. The solid black line is
    the reference case, including only a random dipole noise component
    in H+V of 6~nrad. The dotted and dashed lines are the results for
    \eighthtp, including the same random dipole noise component.}
  \label{fig:8thsim}
\end{figure*}

\begin{figure*}
  \begin{tabular}{cc}
    No excitation & 8th, H+V \\
    \includegraphics[width=\fmawidth]{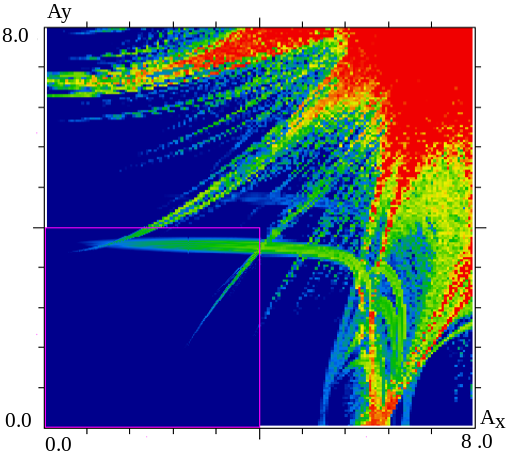} &
    \includegraphics[width=\fmawidth]{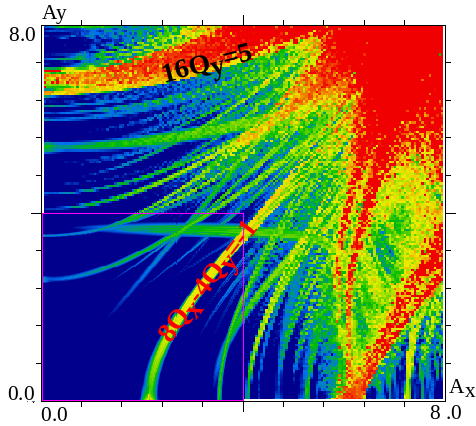}
  \end{tabular}
  \caption{FMA in transverse amplitude space without excitation (left)
    and for \eighthtp\ (right) based on the 2017 injection optics with
    no lattice errors and tunes (62.27, 60.295). The excitation
    amplitude is 96~nrad in both planes. The $16 Q_y$ and
    \mbox{$8 Q_x - 4 Q_y$} resonances are excited.}
  \label{fig:8th2017fmaamp}
\end{figure*}

\begin{figure*}
  \begin{tabular}{c}
  \includegraphics[height=\onethirdheight]{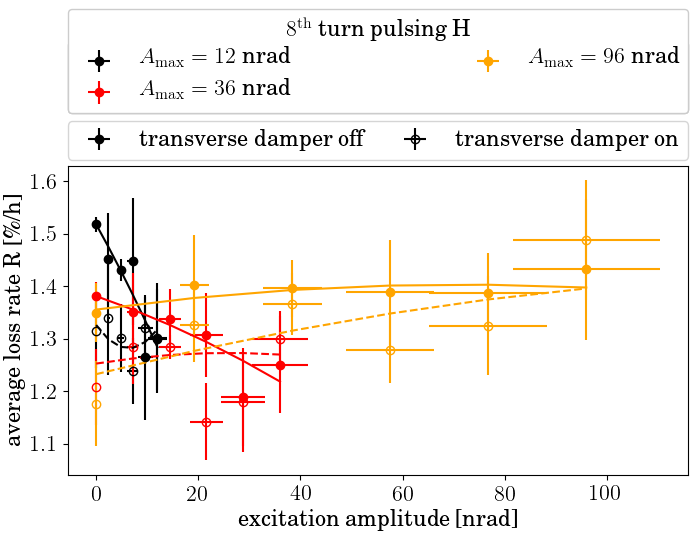} \\
  \includegraphics[height=\onethirdheight]{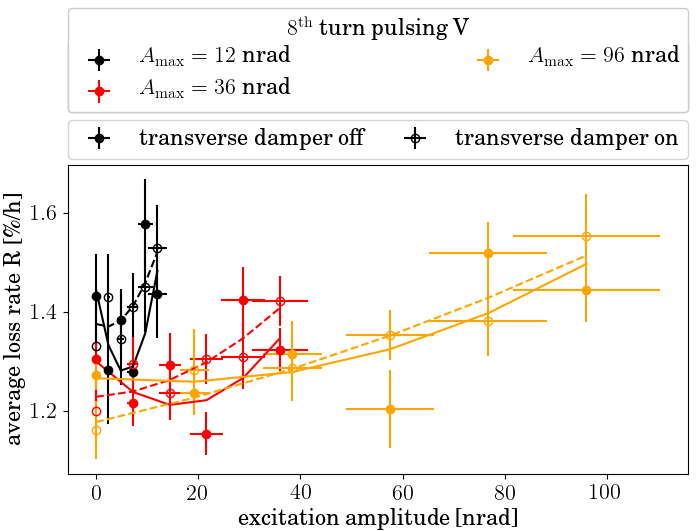} \\
  \includegraphics[height=\onethirdheight]{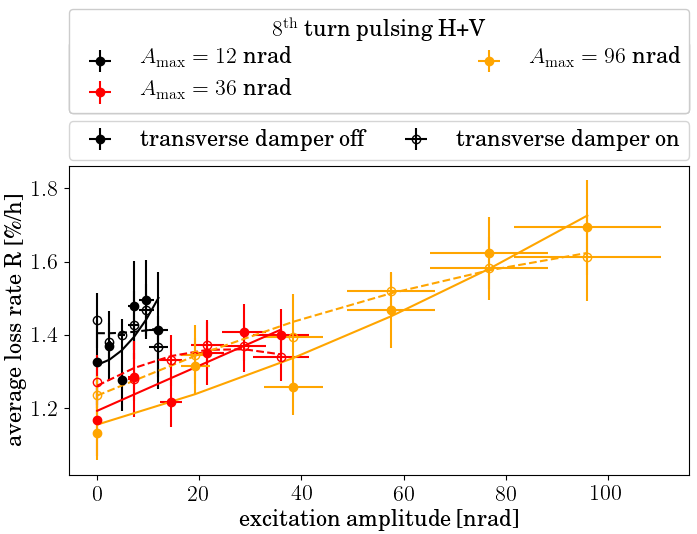} \\
  \end{tabular}
  \caption{Measured loss rates vs.\ excitation amplitude for
    \eighthtp\ in H only (left), in V only (center) and in H+V (right)
    during the 2017 experiment. Uncertainties are statistical. The
    differences between the three data sets within each excitation
    mode (black, red, and yellow curves) provide an estimate of the
    systematic errors.}
  \label{fig:8thexploss2017}
\end{figure*}

\begin{figure*}
  \begin{tabular}{cc}
    Reference bunch, no excitation & Excited bunch, 8th, H \\
    \includegraphics[width=\bsrtwidth]{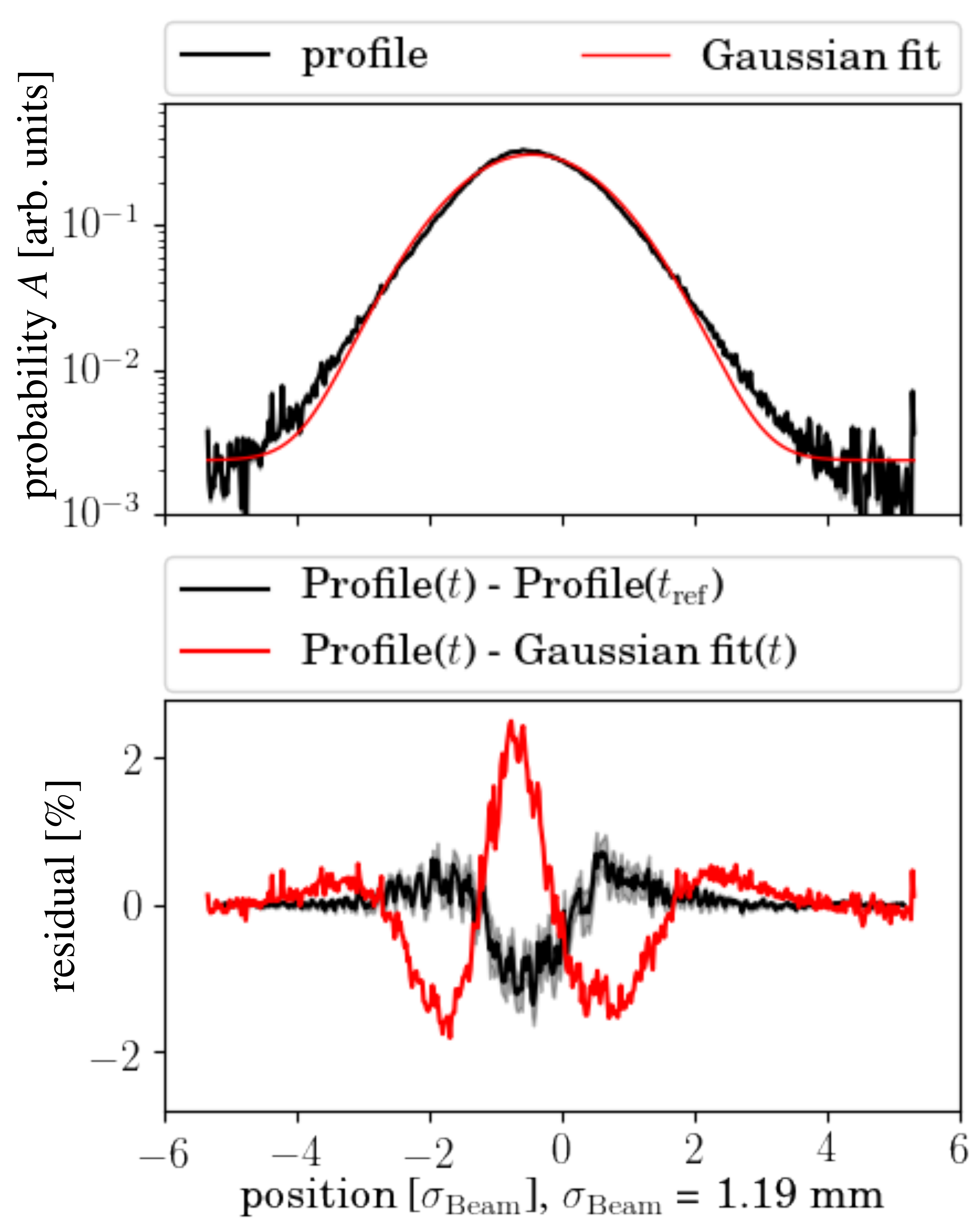} &
    \includegraphics[width=\bsrtwidth]{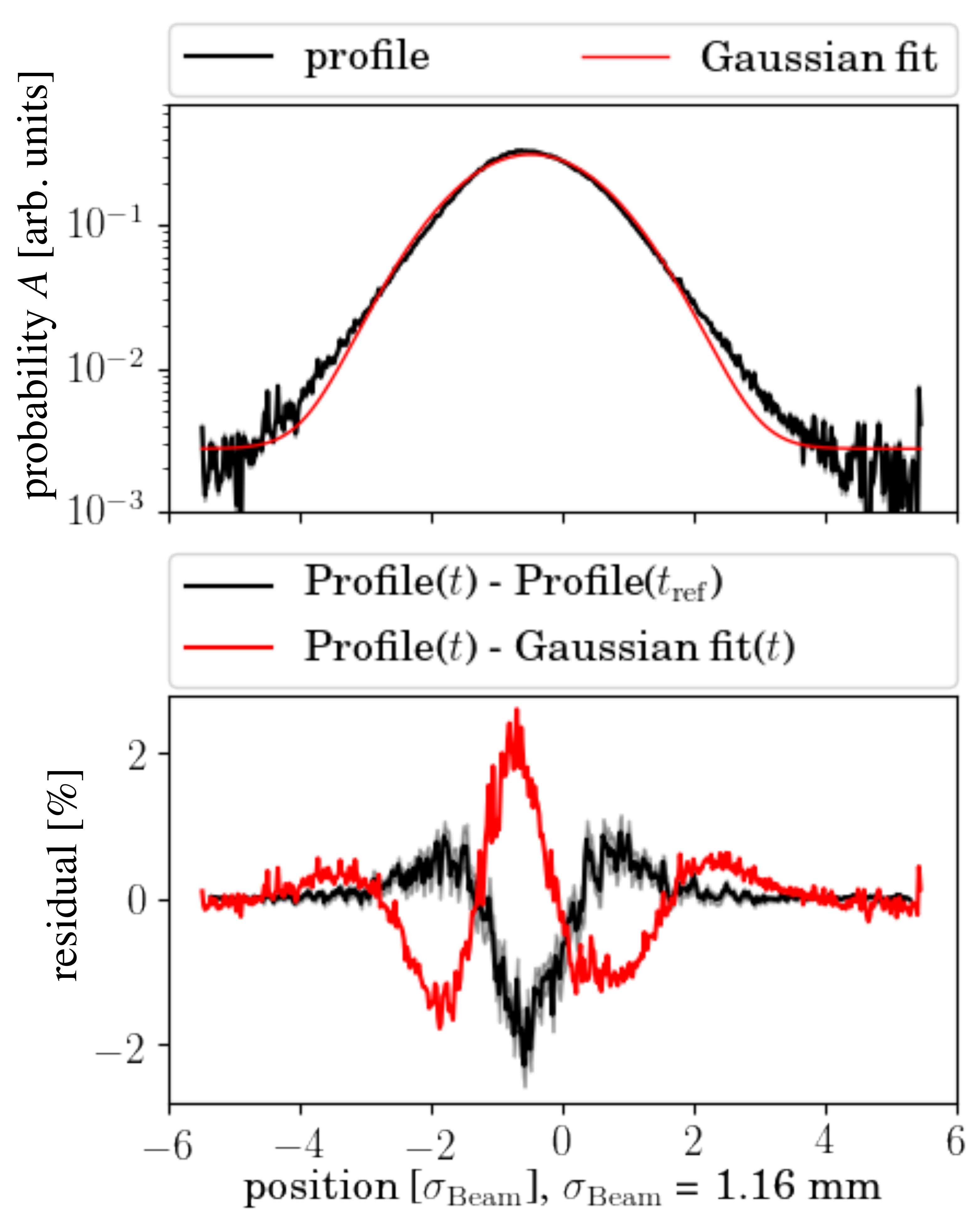} \\
   \end{tabular}
   \caption{Horizontal beam profiles measured with the Beam
     Synchrotron Radiation Telescope (BSRT) during the 2017
     experiments. The profiles are taken at the end of the \eighthtp\
     in H. For these bunches, the transverse damper was not
     active. The data are presented in the same way as in
     Fig.~\ref{fig:10thexpprof}. The profile changes in the bunch
     affected by the maximum excitation (right) are larger than those
     displayed by the control bunch (left).}
   \label{fig:8thexpprof}
\end{figure*}

Excitation patterns that have little effect on the beam core and a
large effect on the halo are good candidates for extending the range
of hollow electron lens operation. In principle, this is possible
because of the highly nonlinear fields generated by the HEL at the
transverse locations of the beam halo and the small fields at the beam
core. For this reason, the \eighthtp\ pattern was chosen for the
experimental studies, as simulations indicated much smaller core
effects compared to 7th- and \tenthtp, as discussed in
Section~\ref{sec:pattern}.

Simulations of intensities, emittances, and bunch lengths for
different amplitudes of the 8th-turn excitation are shown in
Fig.~\ref{fig:8thsim}. No significant effects are predicted for
excitation amplitudes up to 192~nrad. Horizontal and vertical
emittances show small changes in growth rate, but without a clear
dependence on the amplitude or plane of the excitation.

The \fma\ in amplitude space is shown in
Fig.~\ref{fig:8th2017fmaamp}. It reveals that the driven resonances
are of high order, mainly $16 Q_y$ and \mbox{$8 Q_x - 4 Q_y$}, and
their effect is therefore expected to be small.

The effects on the beam were tested during the 2017 experiment. The
excitation amplitude was increased to a maximum of 96~nrad. The loss
rates as a function of excitation amplitude are presented in
Fig.~\ref{fig:8thexploss2017}. For H and V excitations, no significant
increase of the loss rate was observed. A weak increase with amplitude
was seen in the case of H+V excitation. Observed and predicted loss
rates were roughly of the same magnitude (about 0.1--0.4\%/h).

Changes in beam distributions were also detected
(Fig.~\ref{fig:8thexpprof}). The horizontal distribution responded to
the H excitation with a slight depletion of the core and a
corresponding increase around $2\sigma$. For an H+V excitation, one
would expect a similar distribution change. However, this change could
not be detected directly, because, in this case, the emittance of the
control bunches was much smaller (\q{1.8}{\mu m}) than the emittance
of the excited bunches (\q{2.6}{\mu m}), and intra-beam scattering
masked the effects of the excitation~\cite{resexmd2017}. In the
vertical plane, beam distributions and emittances were not affected
by the resonant excitation.

\subsection{Random excitation}
\label{sec:simexran}

\begin{figure*}
  \begin{tabular}{cc}
    No excitation & Random excitation, H+V \\
    \includegraphics[width=\fmawidth]{2017injnocolc15o+19_6noerru_dp0_amp.png} &
    \includegraphics[width=\fmawidth]{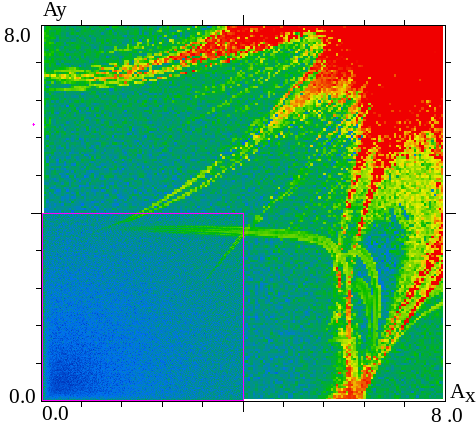} \\
  \end{tabular}
  \caption{FMA in transverse amplitude space without excitation (left)
    and with a random 1-nrad H+V excitation (right), based on the 2017
    injection optics with no lattice errors and tunes (62.27,
    60.295).}
  \label{fig:ran2017fmaamp}
\end{figure*}

\begin{figure*}
  \begin{tabular}{cc}
    \includegraphics[width=\halfwidth]{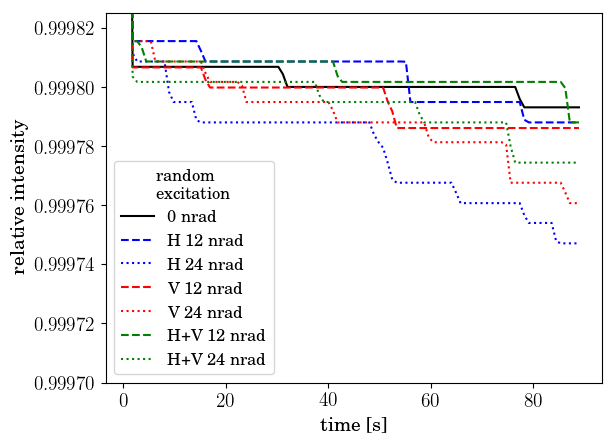} &
    \includegraphics[width=\halfwidth]{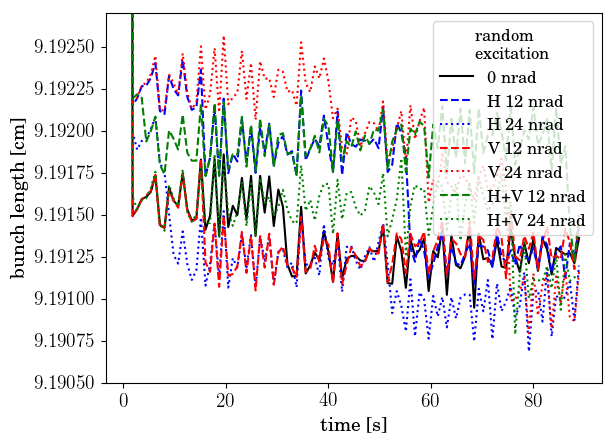}\\
    \includegraphics[width=\halfwidth]{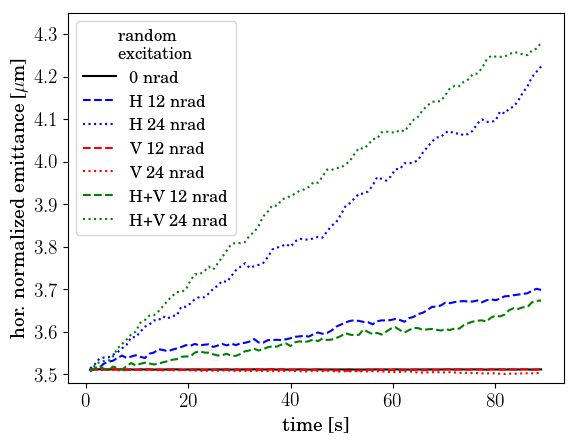} &
    \includegraphics[width=\halfwidth]{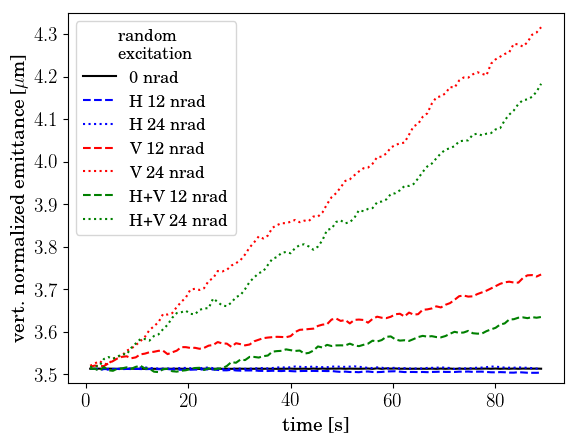}\\
  \end{tabular}
  \caption{Simulations (distribution tracking) based on the 2017
    injection optics with standard lattice errors and
    $(Q_x, Q_y) = (62.27, 60.295)$: relative bunch intensity (top
    left), bunch length (top right), horizontal emittance (bottom
    left), and vertical emittance (bottom right). The solid black line
    indicates the reference case with no excitation. The dashed and
    dotted lines are the results of random excitations (H, V, or H+V)
    with amplitudes 12~nrad and 24~nrad.}
  \label{fig:ransim}
\end{figure*}

\begin{figure*}
  \begin{tabular}{ccc}
    \multicolumn{3}{c}{Random excitation, V} \\
    \includegraphics[width=\thirdwidth]{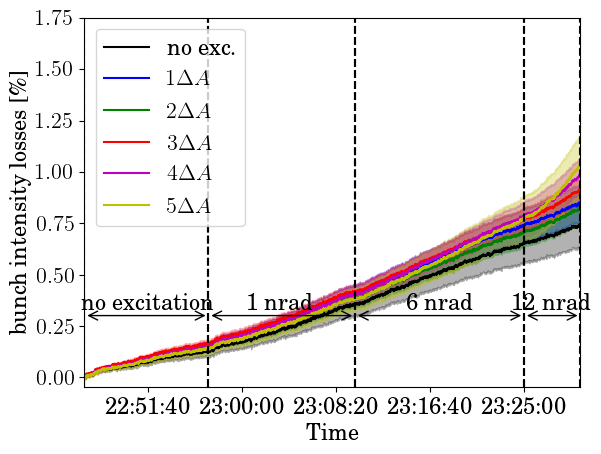} &
    \includegraphics[width=\thirdwidth]{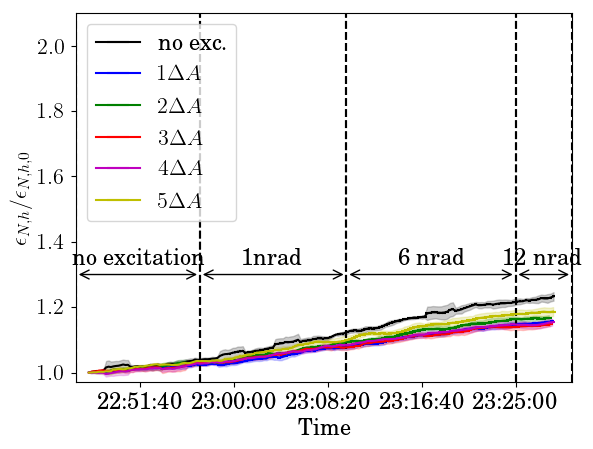} &
    \includegraphics[width=\thirdwidth]{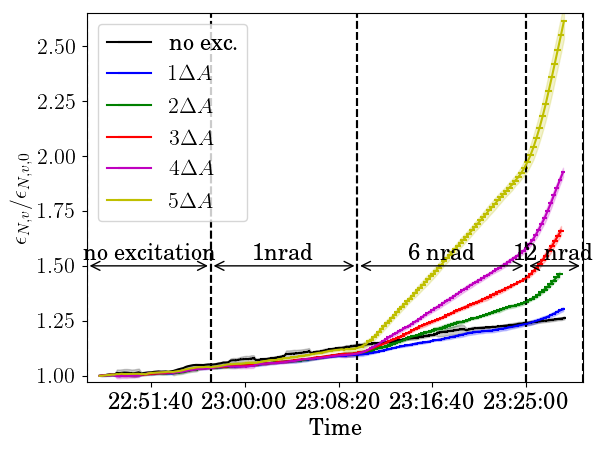} \\
    \multicolumn{3}{c}{Random excitation, H+V} \\
    \includegraphics[width=\thirdwidth]{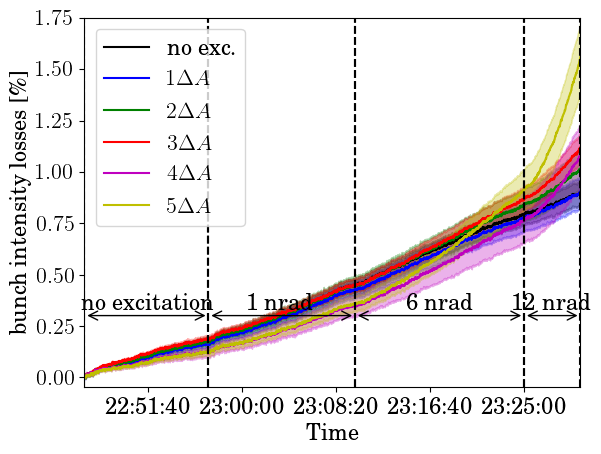} &
    \includegraphics[width=\thirdwidth]{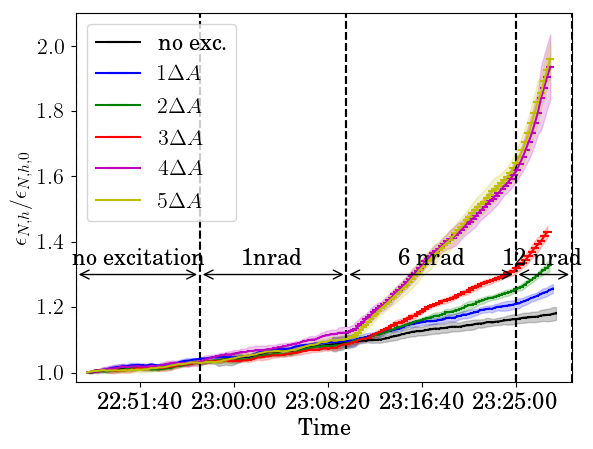} &
    \includegraphics[width=\thirdwidth]{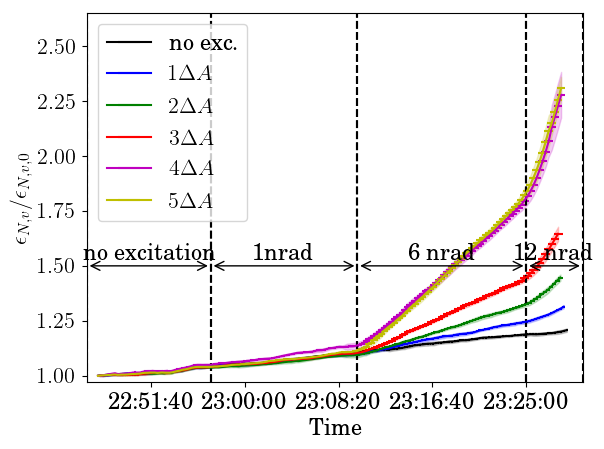}\\
  \end{tabular}
  \caption{Measured effects of the random excitation (in V, top row;
    and H+V, bottom row) during the 2017 experiment: relative
    intensity losses (left), relative horizontal emittance (center),
    and relative vertical emittance (right).}
  \label{fig:ranexp}
\end{figure*}

\begin{figure*}
  \begin{tabular}{cc}
    Reference bunch, no excitation & Excited bunch, random excitation, V \\
    \includegraphics[width=\bsrtwidth]{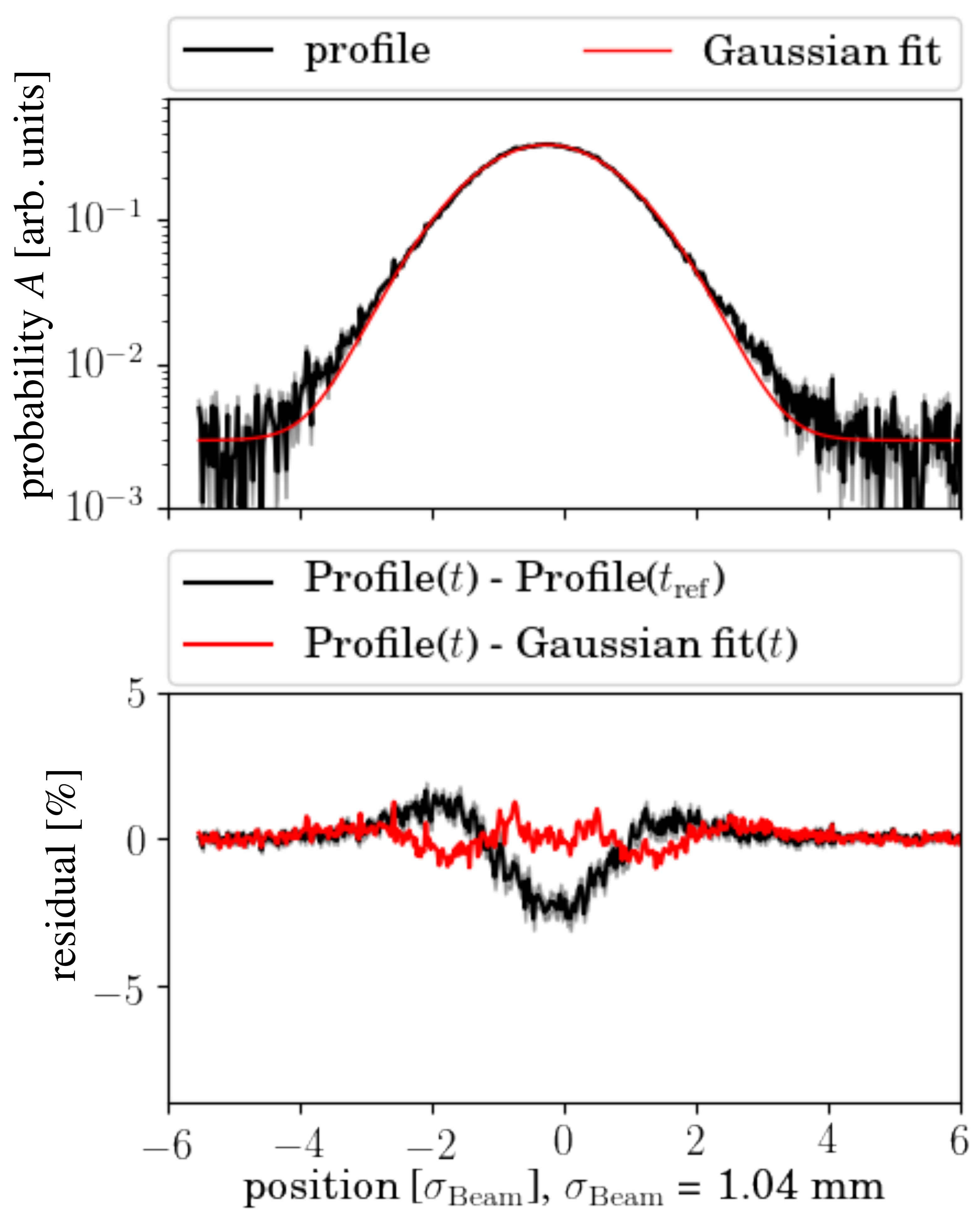} &
    \includegraphics[width=\bsrtwidth]{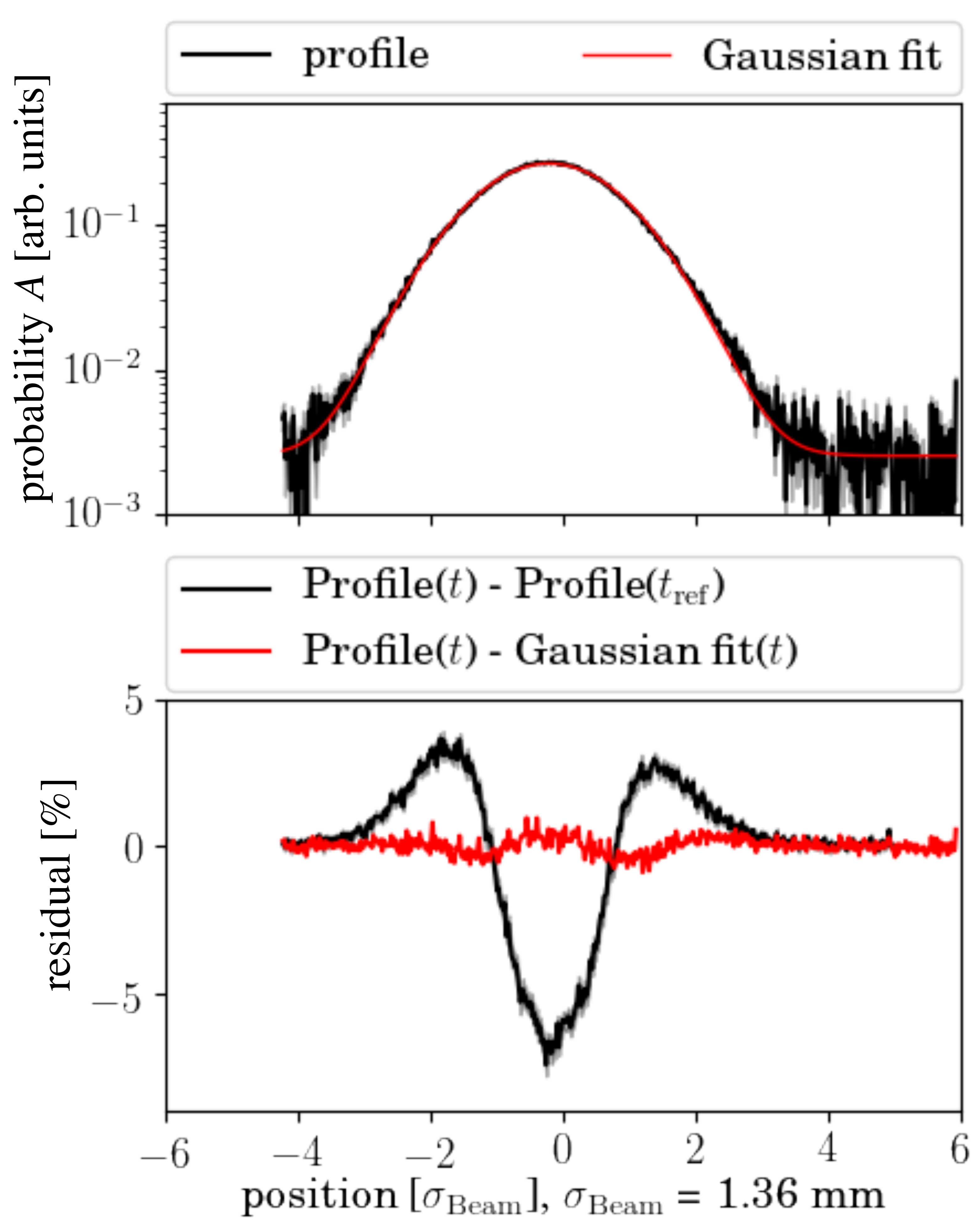} \\
  \end{tabular}
  \caption{Vertical beam profiles measured with the Beam Synchrotron
    Radiation Telescope (BSRT) during the 2017 experiments. The
    profiles are taken at the end of the random excitation in V. For
    these bunches, the transverse damper was not active. The data are
    presented in the same way as in Fig.~\ref{fig:10thexpprof}. The
    distribution changes in the bunch experiencing the maximum
    excitation (right) were larger than those in the reference bunch
    (left). In both cases, distributions retained a Gaussian shape.}
  \label{fig:ranexpprof}
\end{figure*}

\begin{figure*}
  \begin{tabular}{c}
  \includegraphics[height=\onethirdheight]{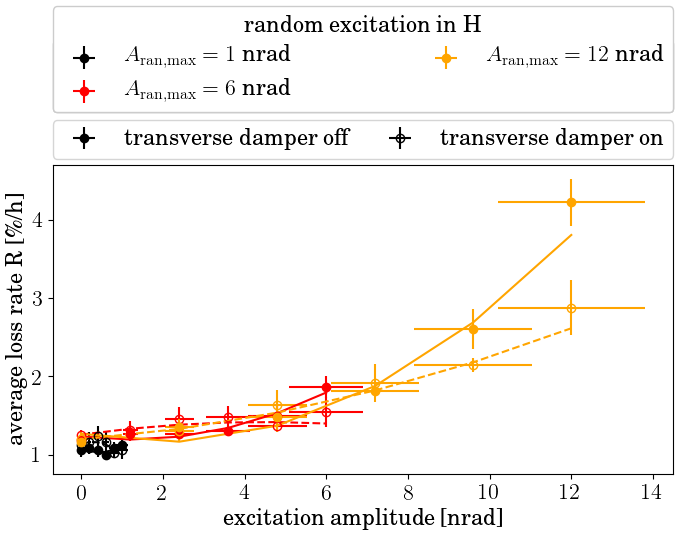} \\
  \includegraphics[height=\onethirdheight]{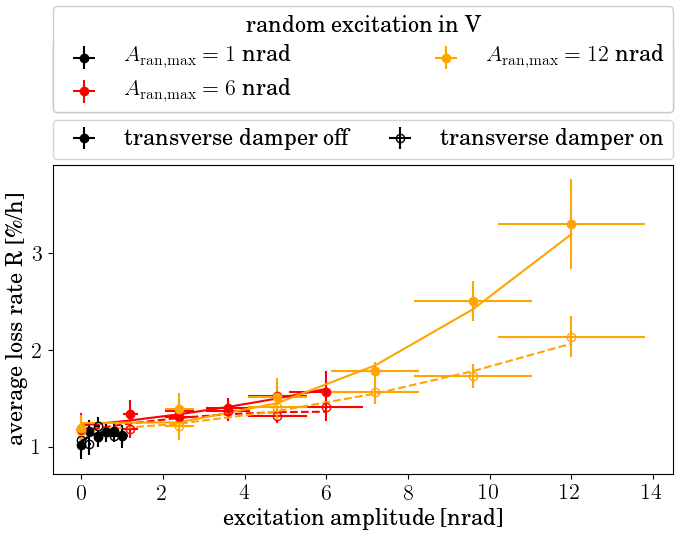} \\
  \includegraphics[height=\onethirdheight]{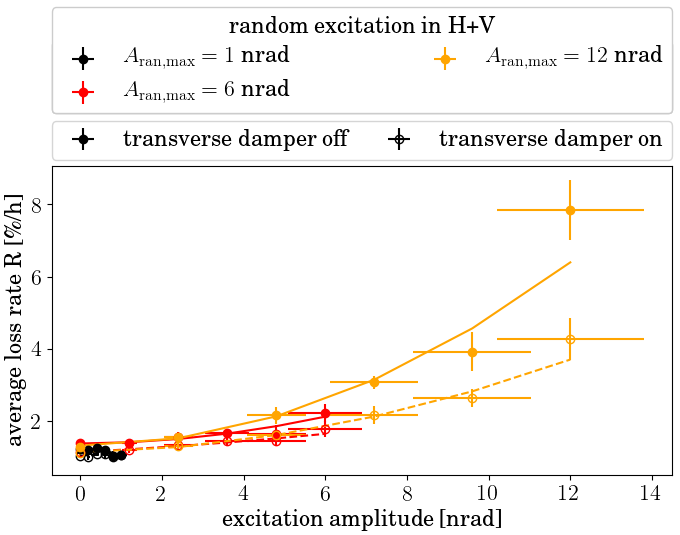} \\
  \end{tabular}
  \caption{Measured loss rates as a function of amplitude for random
    excitations in H only (left), V only (center), and H+V
    (right). For each of the three excitation modes, three consecutive
    data sets were taken (black, red, and yellow), with increasing
    maximum amplitude. Data were taken simultaneously with no
    transverse damper on some bunches (filled circles and solid lines)
    and with the damper active on other bunches (empty circles and
    dashed lines). The lines represent empirical second-order
    polynomial fits.}
  \label{fig:ranexplossplane}
\end{figure*}

\begin{figure*}
  \centering
  \includegraphics[width=\twothirdswidth]{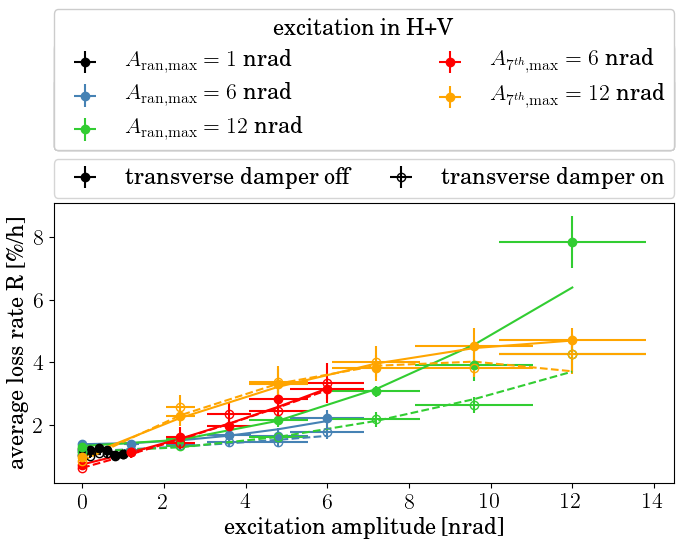}
  \caption{Comparison of measured loss rates vs.\ excitation amplitude
    between random and \seventhtp\ in H+V, obtained during the 2017
    experiment.}
  \label{fig:ranexploss}
\end{figure*}

The random excitation is a powerful and flexible halo-cleaning
pattern, but it is also potentially harmful for the proton beam
core. The effects of random noise in various machines have been
studied extensively in the past (see, for instance,
Refs.~\cite{noise_lebedev_ssc, noise_alexahin_lhc, noise_2007_ohmi,
  noise_2014_ohmi, md_noise_bbLHC}).

A random pattern can be employed at full strength or added as a
modulation to a continuous mode of operation (depending on the
parameter $a$ in Eq.~\ref{eq:random}). During the 2017 experiment, we
tested the effects of the random excitation at full strength ($a = 1$)
and compared it with the resonant excitations.

The random excitation excites practically all resonant frequencies in
the beam.  This can be clearly seen in the FMA example in amplitude
space illustrated in Fig.~\ref{fig:ran2017fmaamp}. Tune diffusion is
increased rather uniformly at all particle amplitudes, and not on
specific resonant lines, as it happens for \kthtp.

The calculated results of distribution tracking, with excitation
amplitudes up to 24~nrad, are shown in
Fig.~\ref{fig:ransim}. They show the following features:
\begin{itemize}
\item negligible losses are observed, independent of the plane of
  excitation;
\item there is no significant change in bunch length;
\item emittance growth is only generated by excitations in the same
  plane (or in both planes);
\item horizontal and vertical emittance growth rates are large and
  comparable in magnitude;
\item when the excitation is applied in both planes, a small increase
  or decrease in the emittance growth rate is observed with respect to
  the cases of separate excitations, an effect likely due to lattice
  coupling;
\item emittance growth takes place at a constant rate, without the
  initial adjustment phase characteristic of \kthtp.
\end{itemize}

The effects of random excitation were measured during the 2017
experiments. Increasing emittance growth with excitation amplitude was
recorded only in the plane of excitation. Horizontal and vertical
emittance growth rates for a given amplitude were comparable. As an
example, the relative horizontal and vertical emittances during V and
H+V pulsing are illustrated in Fig.~\ref{fig:ranexp}. In addition, the
excitations translated into constant growth rates. In particular,
there was no adjustment phase followed by an equilibrium state, as
observed during 7th- and \tenthtp. The qualitative behavior of
emittances predicted in simulations was therefore confirmed.
Calculations could also reproduce the magnitude of the measured
effects. For instance, at 6~nrad, the calculated emittance growth
rates were 3.6\%/min in the horizontal plane and 2.7\%/min vertically,
whereas in experiments we measured 5.0\%/min.

The changes in beam distributions were also directly detected in the
BSRT profiles. As an example, Fig.~\ref{fig:ranexpprof} shows the
profiles induced by a vertical random excitation, compared to those of
a reference bunch. In this case, the distribution widens, but it
retains a Gaussian form, in contrast to the case of \tenthtp, when it
assumed a non-Gaussian shape (see Fig.~\ref{fig:10thexpprof}, for
instance).

The measured loss rates are presented in
Fig.~\ref{fig:ranexplossplane} as a function of random excitation
amplitude in H, V, and H+V planes.  Losses increased quadratically
with excitation amplitude, contributing approximately 3\%/h at
12~nrad. In simulations, the qualitative behavior was the same;
however, the predicted magnitude was much lower (0.1\%/h at
12~nrad). Moreover, the measured effect of the combined H+V excitation
was approximately the sum of H and V separately.

The systematic effects on losses during random excitations were
smaller than during resonant pulsing, as shown by the agreement
between the 3 data sets (black, red, yellow) within each of the
excitation modes in Fig.~\ref{fig:ranexplossplane}. This fact may be
an indication of the random excitation affecting the whole beam,
whereas, in the cases under study, resonant pulsing patterns excited
specific amplitude regions and were therefore more susceptible to
their population and to the order in which experiments were
performed. Another indication comes from the direct comparison of loss
rates vs.\ excitation amplitude for random and \seventhtp\
(Fig.~\ref{fig:ranexploss}). The first time it is applied (red
points), the 7th-turn excitation grows quadratically with amplitude,
whereas the second time it shows signs of saturation (yellow). In any
case, although the dynamics of their action may be different, these
two pulsing patterns were the most powerful, generating losses at the
level of several percent per hour at an amplitude of 12~nrad.

Another feature of random excitation was that the loss rates were
reduced by about a factor~2 when the transverse damping system was
active (dashed vs.\ solid lines in
Fig.~\ref{fig:ranexplossplane}). This was true for all three modes of
excitation (H, V, and H+V). On the other hand, for the resonant
pulsing modes, the transverse damper seemed to have a negligible
effect. This observation is discussed in more detail in
Section~\ref{sec:damp}.

\subsection{Effect of the transverse damper}
\label{sec:damp}

\begin{figure*}
  \begin{tabular}{ccc}
    \multicolumn{3}{c}{\emph{Transverse damper off}} \\
    Random excitation, V (2017) & \seventhtp, V (2017) & \tenthtp, V
                                                         (2016) \\
    \includegraphics[width=\thirdwidth]{2017_emitv_avg_rel_vran_no_damper.png}&
    \includegraphics[width=\thirdwidth]{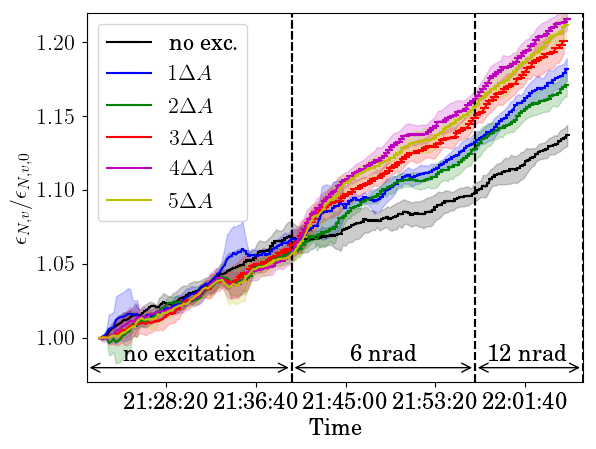} &
    \includegraphics[width=\thirdwidth]{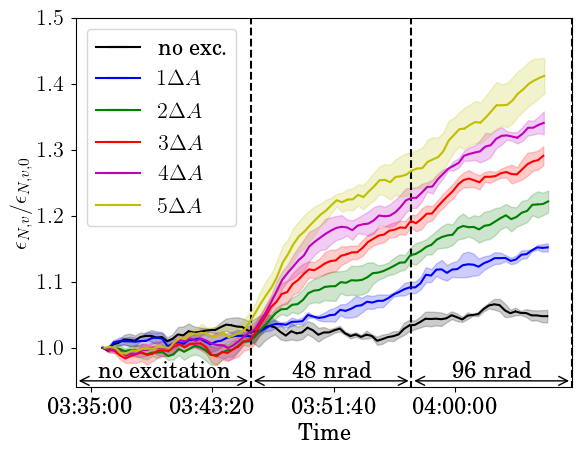}\\
    \multicolumn{3}{c}{\emph{Transverse damper active}} \\
    Random excitation, V (2017) & \seventhtp, V (2017) & \tenthtp, V
                                                         (2016) \\
    \includegraphics[width=\thirdwidth]{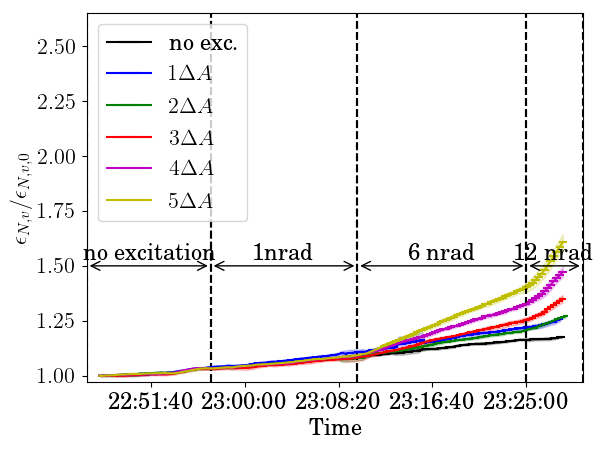} &
    \includegraphics[width=\thirdwidth]{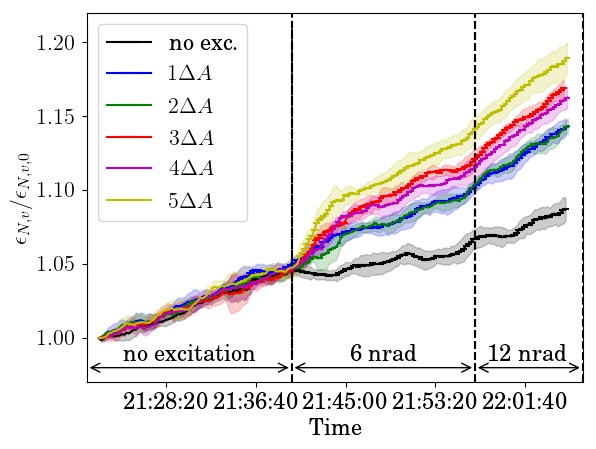} &
    \includegraphics[width=\thirdwidth]{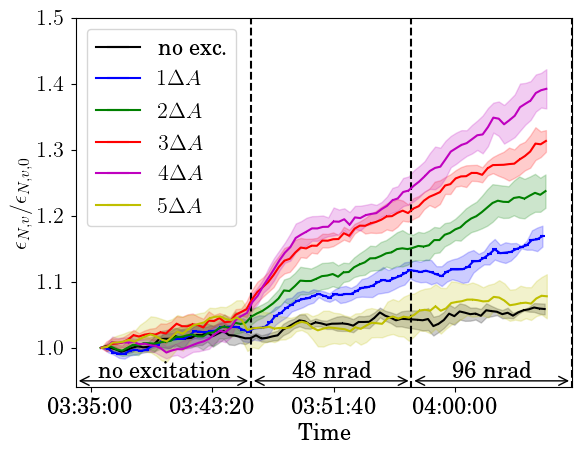} \\
    \end{tabular}
    \caption{Comparison of measured vertical emittances for bunches
      with transverse damper off (top row) and with transverse damper
      on (bottom row), during random V excitations (left), \seventhtp\
      in V (center) and \tenthtp\ in V (right). (Because of a change
      in experimental setup, for the bunches represented by the yellow
      line in the bottom right plot, the excitation was zero instead
      of the maximum value.)}
    \label{fig:damp}
\end{figure*}

During \kthtp\ experiments, the action of the transverse feedback and
damping system (ADT) did not significantly change any of the
observables, namely losses, emittances, and beam distributions. On the
other hand, the ADT considerably reduced any changes of the above
parameters in case of random excitations.

An example is shown in Fig.~\ref{fig:ranexploss}, where losses from
\seventhtp\ in H+V are compared with those measured during random
excitations in H+V. For \seventhtp, the results with and without
damper almost coincide, whereas a significant reduction was observed
for random excitations. As another example, vertical emittances for
7th- and \tenthtp\ in V are compared to those obtained during vertical
random excitations in Fig.~\ref{fig:damp}.

The reason why the transverse damper appeared to be capable of damping
random but not resonant excitations is unknown. In principle, it
fulfills the requirements needed to detect and damp the resulting
oscillations:
\begin{itemize}
\item the closed-orbit distortion caused by the resonant excitations
  was large enough to be detectable by the ADT pickups;
\item in order to detect the orbit distortion, the system compares the
  position of each bunch with its position in the previous turn;
  therefore, an oscillation due to \kthtp, with $k > 1$, should be
  detected and damped;
\item the transverse damper acts on each bunch individually, so it
  should be capable of applying the appropriate corrective kick to
  each group of bunches subject to the same excitation amplitude.
\end{itemize}

We report these observations in an attempt to advance the
understanding of the interaction between resonant excitations and
damping systems. Further studies are obviously needed.

\section{Discussion and conclusions}
\label{sec:sum}

We described the results of numerical simulations and experimental
studies conducted in~2016 and in~2017 on the effects of random and
resonant excitations on proton losses, emittances, and beam
distributions in the Large Hadron Collider.

These studies were motivated in part by the need to assess the effects
of a hollow electron lens (HEL), which is being proposed for active
halo control in HL-LHC. Hollow electron beam collimation was
demonstrated in the Tevatron in continuous
mode~\cite{hel_tevatron_stancari}, i.e., with the same electron beam
current acting on a given bunch every turn. In this case, the unwanted
residual kicks, which arise from profile asymmetries, injection and
extraction bends, solenoid field misalignments, etc. have a negligible
effect on the circulating beam. An electron lens that is pulsed
resonantly can have much stronger effects, and this fact was used in
the Tevatron for abort-gap
cleaning~\cite{hel_tevatron_abortgap_zhang}. The resonant-pulsing
capability (having different currents every turn for a given set of
bunches) will be incorporated in the HL-LHC hollow electron lens to
enhance its range of achievable removal rates. However, due to time
constraints, only a very limited set of experimental tests with
resonant kicks could be done in the Tevatron. More studies are
currently being planned using the electron lenses in the Relativistic
Heavy Ion Collider (RHIC) at Brookhaven National
Laboratory~\cite{Fischer:PRL:2015, Gu:PRAB:2017}.

The effects of resonant excitations depend on several factors,
including machine optics and the distribution of the beam in phase
space and in betatron tunes. For this reason, we decided to study
resonant transverse kicks on proton beams directly in the LHC. We
relied on the flexibility of the LHC transverse feedback and damping
system (ADT) to emulate the nonlinear residual kicks of a hollow
electron lens with transverse dipole kicks, using various excitation
patterns. We focused on the effects on the beam core, namely losses,
emittances and beam distributions, as these determine the magnitude of
the residual fields that can be tolerated. Obviously, the choice of
pulsing patterns to be used in operations depends also on their
effects on the beam halo. These effects are outside the scope of the
present work (and they are challenging to measure directly).

The dipolar component of the residual kick from the electron lens is
approximately 0.5~nrad from the injection/extraction bends and about
20~nrad from profile imperfections in the main overlap region. These
estimates are based on current HL-LHC parameters and on electron-beam
measurements on a test stand. These values can be reduced by improving
cathode emission, solenoid-field uniformity, and transport of the
intense magnetized electron beams.

The effects of the pulsing patterns were evaluated by numerical
tracking simulations and by frequency-map analysis. These calculations
provided insights on the rich nonlinear dynamics of resonant
excitations. Kicking the beam every 7th turn and every 10th turn were
chosen as examples of strong excitations, whereas \eighthtp\ had very
weak consequences.

These patterns were tested in experiments and compared with random
excitations. Bunch filling schemes were devised so that reference
bunches and different excitation amplitudes could be measured at the
same time, including the presence or absence of transverse
feedback. The influence of collective effects was minimized by
reducing the bunch charge. Because these were the first LHC
experiments on resonant excitations, and to reduce turn-around time,
all studies were conducted at injection energy.

The experiments confirmed the relative strength of the pulsing
modes. For instance, random and \seventhtp\ generated losses at the
level of a few percent per hour at 12~nrad, whereas the same level of
losses was obtained at 96~nrad for \tenthtp; even at 96~nrad, losses
from \eighthtp\ did not exceed 0.6\%/h. A similar hierarchy was
observed with respect to emittance growth; in this case, \tenthtp\ was
stronger than \seventhtp. In many cases, simulations underestimated
losses and emittance growth. One important unknown factor was the
magnitude of machine noise and its sources, which had to be
extrapolated from collision to injection energy. Other factors were
collective effects such as intra-beam scattering and electron cloud,
which were minimized in experiments but could not be entirely
avoided. The comparison between \seventhtp\ in 2016 and 2017
emphasized the sensitivity of horizontal and vertical excitations to
working point and tune footprint.

A clear distinction between resonant and random excitations was their
effect on beam distributions. Random noise caused smooth emittance
growth and widening Gaussian distributions. Resonant excitations
generated a fast adjustment of the beam distribution to a new,
non-Gaussian form, followed by a phase of steady evolution. This
behavior was predicted in simulations and was clearly observed in
measured synchrotron-radiation profiles. Potential systematic effects
due to the synchrotron-radiation detection system were mitigated by
directly comparing excited bunches with unaffected bunches used as
controls. Another difference between the two types of excitation was
the repeatability of the strength of the random excitation vs.\ the
dependence of \kthtp\ on the order in which the excitations were
implemented --- later applications being weaker. This confirmed the
hypothesis, emphasized by frequency-map analysis, that random kicks
excite most of the beam, whereas resonant pulsing drives specific
subsets of particles in tune or amplitude space. Finally, the
transverse damping system strongly mitigated losses and emittance
growth generated by random excitations, but had negligible effect on
resonant pulsing patterns.

For active halo control in HL-LHC, resonant pulsing of the electron
lenses, or the addition of a random noise component to continuous
operation, may significantly extend the achievable range of halo
removal rates.  In these studies, we identified candidate excitation
patterns (such as \eighthtp) that preserve the beam core, and their
effects were quantified. As mentioned above, complementary
calculations and experiments are needed to evaluate their efficacy for
halo removal. Because of the complex dynamics, tolerances on residual
kicks vary widely, from a few nanoradians for the most powerful (such
as random at full strength or \seventhtp) to about a hundred
nanoradians for the more benign (such as \eighthtp). The magnitude of
the acceptable kicks depends on the excitation pattern, on machine
lattice, and on the type of application (fast scraping before
collisions vs.\ long-term tail suppression during the course of a
physics fill, for instance). Of course, sensitivity to the
experimental conditions may make resonant patterns less attractive
than a small random noise component added to a constant kick.

These studies are the first systematic investigation of the effects of
resonant excitations on the proton beam in the LHC.  For practical
reasons, the studies were done at injection energy. Because of the
different lattices, working point, noise sources, etc. results at
collision energy may be different and should be investigated.

Because of their flexibility, the experimental methods and modeling
tools developed in this work can be applied more generally to the
investigation of other classes of resonant excitations, beyond pulsed
hollow electron lenses for active halo control.

\begin{acknowledgments}
  There are several people we would like to acknowledge for their
  support, insights, participation in studies, and assistance in data
  collection. In particular, Giulia Pa\-potti offered essential
  support for machine operations and helped prepare the 2016
  experiments. Roderik Bruce and Gianluca Va\-len\-ti\-no made
  important suggestions on the design of the experiments. Dmitry
  Sha\-ti\-lov shared his deep understanding of numerical tracking
  simulations with the \code{lifetrac} code. St\'{e}\-phane
  Far\-toukh, Ric\-car\-do De Maria and Ro\-ge\-lio Tom\'{a}s offered
  invaluable assistance in generating the appropiate accelerator
  lattice models. We wish to thank Sergei Na\-gai\-tsev for the
  insightful discussions on the interaction between the transverse
  damping system and resonant excitations. We are grateful to
  En\-ri\-co Bra\-vin, Georges Trad, and the whole beam
  instrumentation team for their crucial support on
  synchrotron-radiation diagnostics. We thank Ste\-fania
  Pa\-pa\-do\-pou\-lou and Fa\-nou\-ria An\-to\-niou for the fruitful
  collaboration on the analysis of the synchrotron-radiation profiles.

  This manuscript has been authored by Fermi Research Alliance,
  LLC under Contract No.~DE-AC02-07CH11359 with the U.S.\ Department
  of Energy (DOE), Office of Science, Office of High Energy
  Physics. This work was partially supported by the U.S.\ DOE LHC
  Accelerator Research Program (LARP), by the European FP7 HiLumi LHC
  Design Study, Grant Agreement 284404, and by the High Luminosity LHC
  (HL-LHC) Project.
\end{acknowledgments}

% bibliography
\bibliography{resex}

\end{document}